\documentclass[12pt]{article}
\usepackage{enumerate}
\usepackage{bm}
\usepackage{bmpsize}
\usepackage{color}
\usepackage{subfigure}
\usepackage{upquote}
\usepackage{url} 
\usepackage{booktabs}
\usepackage{graphicx,psfrag,epsf}
\usepackage{epstopdf}
\usepackage{amsmath}
\usepackage{amssymb}
\usepackage[maxfloats=36]{morefloats}
\usepackage{float}
\usepackage[english]{babel}
\usepackage[titletoc,title]{appendix}
\usepackage{natbib}
\usepackage{upquote}
\usepackage{indentfirst}

\usepackage{bmpsize}
\usepackage{color,soul} 
\usepackage{xcolor}
\usepackage{amssymb}
\usepackage{comment}
\usepackage[colorlinks]{hyperref}
\hypersetup{
  colorlinks=true,
  citecolor=blue,
  linkcolor=red,
  }
\usepackage{authblk}
\usepackage{caption} 
\usepackage{multicol}
\usepackage{multirow}

\begin{document}

\title{\bf Robust regression based on shrinkage estimators}

\author[1]{Elisa Cabana\thanks{This research was partially supported by Spanish Ministry grant ECO2015-66593-P.}}
\author[2]{Rosa E. Lillo}
\author[3]{Henry Laniado}

\affil[1]{Statistics Department, Universidad Carlos III de Madrid, Spain.}
\affil[2]{Statistics Department and UC3M-Santander Big Data Institute, Universidad Carlos III de Madrid, Spain.}
\affil[3]{Mathematical Sciences Department,  Universidad EAFIT, Medell\'in, Colombia.}
\date{}                     
\setcounter{Maxaffil}{0}
\renewcommand\Affilfont{\itshape\small}

\maketitle

\bigskip

\begin{abstract}
A robust estimator is proposed for the parameters that characterize the linear regression problem. It is based on the notion of shrinkages, often used in Finance and previously studied for outlier detection in multivariate data. A thorough simulation study is conducted to investigate: the efficiency with normal and heavy-tailed errors, the robustness under contamination, the computational times, the affine equivariance and breakdown value of the regression estimator. Two classical data-sets often used in the literature and a real socio-economic data-set about the Living Environment Deprivation of areas in Liverpool (UK), are studied. The results from the simulations and the real data examples show the advantages of the proposed robust estimator in regression.
\end{abstract}

{\it Keywords:}  robust regression, robust Mahalanobis distance, shrinkage estimator, outliers.
\vfill
\hfill {}

\newpage


\maketitle

\section{Introduction}

Linear regression problems are widely used in numerous fields. The diversity of data for which the model is used poses a problem since not all available methods work well for high dimension, high sample size, not all are sufficiently resistant to the presence of anomalous values, and are computationally feasible at the same time. Consider the linear regression model:
\begin{equation}\label{model}
y_i=\alpha  +  \mathbf{x}_i^t \boldsymbol{\beta} + \epsilon_i
\end{equation}

\noindent
for $i=1,...,n$, where $n$ is the sample size, $\alpha$ is the unknown intercept, $\boldsymbol{\beta}$ is the unknown $(p \times 1)$ vector of regression parameters, and the error terms $\epsilon_i$ are i.i.d and also independent from the $p$-dimensional explanatory variables $\mathbf{x}_i$ (often also called regressor variables or carriers). The classical approach to estimate the parameters of the model is the ordinary least squares (OLS) estimator of Gauss and Legendre, which minimizes the sum of squared residuals:
\begin{equation}\label{betaols}
\hat{\boldsymbol{\beta}}_{OLS}=\underset{\boldsymbol{\beta}}{argmin} \sum_{i=1}^n (y_i - \mathbf{x}_i^t \boldsymbol{\beta})^2
\end{equation}

The problem with OLS is that a single unusual observation can have a large impact on the estimate. Through all these past three decades there have been different approaches attempting the robustification of the procedure, although there is no consensus that establishes which method is recommended in practical situations. OLS estimator can be expressed as follows. Denote the joint variable of the response and carriers as $\mathbf{z=(x,y)}$. Denote the location of $\mathbf{z}$ by $\boldsymbol{\mu}$ and the scatter matrix by $\Sigma$. Partitioning $\boldsymbol{\mu}$ and $\Sigma$ yields the notation:
\begin{equation}\label{mucov1}
\begin{matrix}
\boldsymbol{\mu}=\begin{pmatrix}
\boldsymbol{\mu}_x\\ 
\mu_y
\end{pmatrix}& , & \Sigma=\begin{pmatrix}
 \Sigma_{xx} & \Sigma_{xy} \\ 
\Sigma_{yx} & \Sigma_{yy}
\end{pmatrix}
\end{matrix} 
\end{equation}

Traditionally they are estimated by the empirical mean $\hat{\boldsymbol{\mu}}$ and the empirical covariance matrix $\hat{\Sigma}$. OLS estimators of $\boldsymbol{\beta}$ and the intercept $\alpha$ can be written as functions of the components of $\boldsymbol{\hat{\mu}}$ and $\hat{\Sigma}$, namely
\begin{align}\label{parameters1}
\hat{\boldsymbol{\beta}} &= \hat{\Sigma}_{xx}^{-1}\hat{\Sigma}_{xy} , \quad  \quad
\hat{\alpha} = \hat{\mu}_{y} -\hat{\boldsymbol{\beta}} ^t \hat{\boldsymbol{\mu}}_{x}  
\end{align}

The drawback is that the classical sample estimators are sensitive to the presence of outliers. Instead, robust estimators should be used. The contribution of this paper is to propose robust estimators based on shrinkage to be used in Equation \ref{parameters1} for estimating the regression parameters (a similar idea can be seen in \cite{maronna1986robust} and \cite{croux2003bounded}. These estimators based on shrinkage have shown advantages when they were used for defining a robust Mahalanobis distance to detect outliers in the multivariate space \citep{cabana2019multivariate} and in the present paper, the performance in linear regression is studied, through simulations and real data examples. The notion of shrinkage is used in Finance and Portfolio optimization, and it provides a trade-off between low bias and low variance (\cite{ledoit2003honey}, \cite{ledoit2003improved}, \cite{ledoit2004well}, \cite{demiguel2013size}), and in case of covariance matrices, well-conditioned estimates are obtained, a fact that is of relevance when inversion of the matrix is at stake, as is the case now.

Furthermore, a real socio-economic example that explains the Living Environment Deprivation (LED) index of areas in Liverpool (UK) through remote sensing data, is studied.  The data was previously used in \cite{arribas2017remote} where two machine learning approaches were  investigated in this context: Random Forest (RF) and Gradient Boost Regressor (GBR). In this paper we study the proposed robust regression approach with the LED index data and found out that it provides an improvement of the cross-validated $R^2$ and mean squared error with respect to classical OLS and both machine learning techniques RF and GBR, while maintaining the advantage of interpretability, which is a weakness that RF and GBR have.

The paper is organized as follows. Section \ref{review} shows a state-of-the-art review of the most used methods for robust regression in the literature. In Section \ref{our}, the alternative robust method based on shrinkage is proposed. The approach is compared with the others by means of simulations. The description of the simulation scenarios is shown in Section \ref{sim}. In Section \ref{seceff} the efficiency is studied with normal errors and heavy-tailed distributed erros. In Section \ref{secrob}, the robustness and the computational performance are investigated in presence of contamination. Section \ref{seceq} shows the equivariance property studied by means of simulations and the breakdown value is shown in Section \ref{secbdp}. On the other hand, real data examples are considered in Section \ref{realex}. Finally, in Section \ref{final} some conclusions are provided.

\section{State of the art}\label{review}

The efficiency and breakdown point (bdp) are two traditionally used criteria to compare the existing robust methodologies. The first one because OLS has the smallest variance among unbiased estimates when the errors are normally distributed and there are no outliers. This means that, in this scenario, OLS has maximum efficiency. Thus, the \textit{relative efficiency} of the robust estimate compared to OLS when the error distribution is exactly normal and the data is clean, is often considered as a measure to study the performance of the methods and to compare them with each other. The bdp measures the proportion of outliers an estimate can tolerate. Usually, the definition of \textit{finite sample bdp} is used \citep{donoho1983notion}. Given any sample $\mathbf{z}=(\mathbf{z}_1,...,\mathbf{z}_n)$, with $\mathbf{z}_i=(\mathbf{x}_i,y_i)$, where $\mathbf{x}_i$ is of dimension $1 \times p$, for all $i=1,...,n$, denote by $T(\mathbf{z})$ an estimate of the parameter $\boldsymbol{\beta}$. Let $\widetilde{\mathbf{z}}$ be the corrupted sample where any $q$ of the original points of $\mathbf{z}$ are replaced by arbitrary outliers. Then the finite sample bdp $\gamma^*$ is defined as:
\begin{equation}
\gamma^*(T,\mathbf{z})=\underset{1\leq q \leq n}{min }  \{ \frac{q}{n} : \underset{\widetilde{z}}{sup } \ || T(\widetilde{\mathbf{z}})- T(\mathbf{z}) || =\infty \}
\end{equation}
\noindent
where $|| \cdot ||$ is the Euclidean norm. The asymptotic bdp is understood as the limit of the finite sample bdp when $n$ goes to infinity. Intuitively, the maximum possible asymptotic bdp is $1/2$ because if more than half of the observations are contaminated, it is not possible to distinguish between the background data and the contamination \citep{leroy1987robust}. OLS has a finite sample bdp of $1/n$ and asymptotic bdp of $0$.

A first proposal of a robust estimate in regression came from \cite{edgeworth1887observations} who proposed to replace the squared residuals in the definition of Equation \ref{betaols} by their absolute value. It was called Least Absolute Deviation (LAD) or $L_1$ estimate and it was more resistant than OLS against outliers in the response variable $y$, but still couldn't resist outlying values in the carriers. These kind of outliers are called \textit{leverage points}, which may have a large effect on the fit. Thus, the finite sample bdp of LAD is $1/n$.

The next idea was made by \cite{huber1964robust} (also see \cite{huber1973robust} and \cite{huberrobust}) who proposed to replace the least-square criterion by a robust loss function $\rho(\cdot)$ of the residuals. It was called M-estimator, which was more efficient than LAD. However, the finite sample bdp of both LAD and M tend to 0, because of the possibility of leverage points \citep{maronna2006robust}. Besides, the method implies one first decision: which loss function $\rho$ should be used. Huber's loss or the Tukey's bisquare functions are common choices, but there are no rules for which should be selected when we are dealing with real data. Furthermore, they depend on a constant that determines the efficiency of the estimator, and this might be a problem as well in practice. Due to the vulnerability of M-estimators, the generalized M-estimators (also called GM-estimators) were proposed, and the problem of recognizing leverage points was solved, but it could not distinguish between ``good'' and ``bad'' leverage points, and the bdp decreases as the dimension $p$ of the data increases.

\cite{siegel1982robust} proposed a near $50\%$ bdp technique, the Least Median of Squares (LMS), which minimizes the median of the squared residuals. However, the procedure had a disadvantage in the order of convergence (\cite{rousseeuw1982least}, \cite{rousseeuw1993alternatives}). Another approach was proposed by \cite{rousseeuw1983multivariate}, called Least Trimmed Squares (LTS) and it consisted on minimizing the sum of the $h$ ordered squared residuals, where $h$ is the proportion of trimming. Usually $h=n/2 + 1$ results in a bdp of $50\%$ and better convergence rate than LMS. The problem is LTS suffers in terms of low efficiency relative to OLS \citep{stromberg2000least}. 

Robust regression by means of S-estimator came by hands of \cite{rousseeuw1984robust}. The method has greater asymptotic efficiency than LTS, but depending on the specification of some constants. \cite{croux1994generalized} proposed the generalized S-estimator (GS-estimator) to improve the efficiency, but again there was a constant to define, which depends on $n$ and $p$.

MM-estimators were proposed by \cite{yohai1987high} and consisted in three basic stages. For the initial step, a consistent robust estimate of the regression parameters with high bdp but not necessarily high efficiency, was needed. In practice the typical initial estimators are LMS or S-estimate with Huber or bisquare functions. Playing with the constants necessary for the estimators, MM-estimates can attain high efficiency without affecting its bdp. However the author recognize in \cite{yohai1987high} that if the constant that handles the efficiency is increased, then the estimates get more sensitive to outliers.



\cite{maronna1986robust} and \cite{croux2003bounded} proposed another idea based on using robust estimators in the expression for OLS estimates from Equation \ref{parameters1}. They propose to use the multivariate M-estimators and the S-estimator (method S from now on), respectively.

The robust and efficient weighted least square estimator (REWLSE) was proposed by \cite{gervini2002class}. The method simultaneously achieve maximum bdp and full efficiency under Gaussian errors. The idea is to use hard rejection weights (0 or 1) calculated from an initial robust estimator. The cut-off depends on the distribution of the standardized absolute residuals, and because of these adaptive cut-off, the method is asymptotically equivalent to OLS and hence its full asymptotic efficiency.


In summary, all these least squares alternatives exhibit some drawbacks. Some are robust to outliers in the response, but not resistant to leverage points, or could not distinguish between good or bad leverage. A maximum bdp is difficult to achieve maintaining high efficiency. MM-estimator, method S and REWLSE estimator seem to be the best alternatives because of their high bdp and high asymptotic efficiency. It is important to note that even though some mentioned estimators have high bdp, their computation is challenging specially in case of large data-sets or high dimension. That is why approximate algorithms have to be used, which are usually based on taking a number of subsamples and iterate. This fact translates in worse performance about consistency and bdp than the exact theoretical estimator would have had. It gets worse with the increase of the sample size $n$ or the dimension $p$ of the samples (\cite{stromberg2000least}, \cite{hawkins2002inconsistency}). Furthermore, with all these methods there always have to be a decision of which tuning constant choose, or which function of the residuals use, or which first initial estimator use. The problem becomes complicated with all of these decisions in case of real data.

\section{Shrinkage reweighted regression}\label{our}

In this paper, robust estimators of location and scatter matrix based on the notion of \textit{shrinkage}, are used in Equation \ref{parameters1}. The notion of shrinkage relies on the fact that ``shrinking'' an estimator $\hat{E}$ of a parameter towards a \textit{target estimator} $\hat{T}$, would help to reduce the estimation error because it is a trade-off between a low bias estimator and a low variance estimator. According to \cite{james1961estimation}, under
general conditions, there exists a shrinkage intensity $\eta$, so the resulting shrinkage
estimator would contain less estimation error than $\hat{E}$.
\begin{equation}\label{shrinkgeneral}
\hat{E}_{Sh}=(1-\eta) \hat{E} + \eta \hat{T}
\end{equation}

Let $\mathbf{x} = \{ \mathbf{x}_{\cdot 1} ,...,\mathbf{x}_{\cdot p}  \} $ be the $n \times p$ data matrix with $n$ being the sample size
and $p$ the number of variables. In \cite{cabana2019multivariate}, the shrinkage estimator $\hat{\boldsymbol{\mu}}_{Sh}$ is proposed as a robust estimator of central tendency. 
\begin{equation}\label{shrinkmuMM}
\hat{\boldsymbol{\mu}}_{Sh}=(1- \eta) \hat{\boldsymbol{\mu}}_{MM} + \eta \nu_{\boldsymbol{\mu}} \mathbf{e} 
\end{equation}

\noindent
where $\hat{\boldsymbol{\mu}}_{MM}$ is the multivariate \textit{$L_1-$median}, which is a robust and highly efficient estimator of location (\cite{lopuhaa1991breakdown}, \cite{vardi2000multivariate}, \cite{oja2010multivariate}).  The target
estimator was $\nu_{\boldsymbol{\mu}} \mathbf{e}$ , where $\mathbf{e}$ is the $p$-dimensional vector of ones, analogous as in \cite{demiguel2013size}. The scaling factor $\nu_{\boldsymbol{\mu}}$ and the intensity $\eta$ are obtained minimizing the expected
quadratic loss. The solution can be found in Proposition 2 from \cite{cabana2019multivariate}. On the other hand, the authors also propose an adjusted special comedian matrix $\hat{S}_{Sh}$, based on the classical definition of comedian from \cite{falk1997mad}, and with it a shrinkage estimator for the covariance matrix can be obtained.
\begin{equation}
\hat{S}_{Sh}=2.198\cdot (median((\mathbf{x}_{\cdot j} - (\boldsymbol{\hat{\mu}}_{Sh})_j ) (\mathbf{x}_{\cdot t} - (\boldsymbol{\hat{\mu}}_{Sh})_t ))
\end{equation}

The idea came from the fact that the comedian matrix is a robust alternative for the covariance matrix, but in general it is not positive (semi-)definite \citep{falk1997mad}, and with the shrinkage approach applied to the comedian, a robust and well-conditioned estimate is obtained (\cite{ledoit2003honey}, \cite{ledoit2003improved}, \cite{ledoit2004well}, \cite{demiguel2013size}). The shrinkage estimator will be:
\begin{equation}\label{shrinkS*}
\hat{\Sigma}_{Sh}=(1- \eta) \hat{S}_{Sh} + \eta \nu_{\Sigma} I
\end{equation}

The optimal expression for the parameters $\eta$ and  $\nu_{\Sigma}$ is described in \cite{cabana2019multivariate} in Proposition 3. Furthermore, the authors used the robust estimators of location $\hat{\boldsymbol{\mu}}_{Sh}$ and covariance matrix $\hat{\Sigma}_{Sh}$ based on shrinkage to define a robust Mahalanobis distance that had the ability to discover outliers with high precision in the vast majority of cases in the simulation scenarios studied in the paper, with both gaussian data and with skewed or heavy-tailed distributions. The behavior under correlated and transformed data showed that the approach was approximately affine equivariant. With highly contaminated data it is shown that the method had high breakdown value even in high
dimension. 

In the present paper, the estimation of the regression parameters using these robust estimators based on shrinkage in Equation \ref{parameters1}, is proposed. Consider the joint vector $\mathbf{z=(x,y)}$ with $\boldsymbol{\mu}$ and $\Sigma$ the location and covariance matrix of $\mathbf{z}$ described in Equation \ref{mucov1}. Now let us call the shrinkage estimators $\hat{\boldsymbol{\mu}}_{Sh}$ and $\hat{\Sigma}_{Sh}$ for the location and covariance matrix of $\mathbf{z}$, the \textit{initial shrinkage robust estimators} of central tendency and covariance matrix of $\mathbf{z}$, respectively. Now let us define the associated robust squared Mahalanobis distance for each observation $\mathbf{z}_i$, with $i=1,...,n$:
\begin{equation}\label{eq:primeradist}
d^2(\mathbf{z}_i )=(\mathbf{z}_i-\hat{\boldsymbol{\mu}}_{Sh})^t \hat{\Sigma}_{Sh}^{-1}  (\mathbf{z}_i-\hat{\boldsymbol{\mu}}_{Sh})
\end{equation}


Let $w_i=w(d^2(\mathbf{z}_i))$ be a weight function depending on the robust squared Mahalanobis distance. The second step is to obtain $\hat{\boldsymbol{\mu}}_{Sh}^{SW}$ and $\hat{\Sigma}_{Sh}^{SW}$, the \textit{shrinkage weighted estimator for the mean and covariance matrix:}

\begin{align}\label{eq:locshrinkrewest2}
\hat{\boldsymbol{\mu}}_{Sh}^{SW} &= \frac{  \sum_{i=1}^n w_i  \mathbf{z}_i    }{   \sum_{i=1}^n w_i  }, \quad
\hat{\Sigma}_{Sh}^{SW} = \frac{  \sum_{i=1}^n w_i  (\mathbf{z}_i - \hat{\boldsymbol{\mu}}_{Sh}^{SW}) (\mathbf{z}_i - \hat{\boldsymbol{\mu}}_{Sh}^{SW})^t }{   \sum_{i=1}^n w_i  } 
\end{align}

Based on $\hat{\boldsymbol{\mu}}_{Sh}^{SW}$ and $\hat{\Sigma}_{Sh}^{SW}$ we can obtain $\hat{\boldsymbol{\beta}}^{SW}$ and $\hat{\alpha}^{SW}$ which are initial estimates for the regression parameters. Let us call them \textit{shrinkage weighted (SW) regression estimators}:
\begin{align}
\hat{\boldsymbol{\beta}}^{SW} &= (\hat{\Sigma}_{Sh}^{SW})_{xx}^{-1}(\hat{\Sigma}_{Sh}^{SW})_{xy} , \quad  
\hat{\alpha}^{SW} = (\hat{\boldsymbol{\mu}}_{Sh}^{SW})_{y} -(\hat{\boldsymbol{\beta}}^{SW})^t (\hat{\boldsymbol{\mu}}_{Sh}^{SW})_{x} 
\end{align}\label{eq:betaL}
The SW error's scale estimate is:
$$\hat{\sigma}^{SW} = (\hat{\Sigma}_{Sh}^{SW})_{yy} -(\hat{\boldsymbol{\beta}}^{SW})^t (\hat{\Sigma}_{Sh}^1)_{xx} \hat{\boldsymbol{\beta}}^{SW}$$
The third step is reweighting, taking into consideration the residuals based on the SW regression estimators:
\begin{equation}
r_i^{SW}=y_i - (\hat{\boldsymbol{\beta}}^{SW})^t \mathbf{x}_i - \hat{\alpha}^{SW}
\end{equation}
Define the Mahalanobis distance for the SW residuals:
\begin{equation}
d(r_i^{SW})=((r_i^{SW})^t(\hat{\sigma}^{SW})^{-1} r_i^{SW} )^{1/2}
\end{equation}
Let $wr_i=w(d^2 (r_i^{SW}))$ a weighting function that depends on the Mahalanobis distance of the SW residuals. Define $\mathbf{u}_i=(\mathbf{x}_i^t , 1)^t$ and obtain:
\begin{equation}
\hat{\boldsymbol{\varphi}}^{SR}=((\hat{\boldsymbol{\beta}}^{SR})^t,\hat{\alpha}^{SR})^t = \left( \sum_{i=1}^n wr_i  \mathbf{u}_i \mathbf{u}_i^t   \right )^{-1} \sum_{i=1}^n wr_i  y_i \mathbf{u}_i
\end{equation}
Then,  $\hat{\boldsymbol{\varphi}}^{SR} = \left ( \left (\hat{\boldsymbol{\beta}}^{SR}\right )^t, \hat{\alpha}^{SR}   \right )^t$ are the \textit{shrinkage reweighted (SR) regression estimators}.\\

For the weighting functions the inverse of the squared robust Mahalanobis distance was studied, but the indicator function in both cases (as in \cite{rousseeuw2004robust}) had improved performance. The first weight function  is $w_i=w(d^2(\mathbf{z}_i))=I(d^2(\mathbf{z}_i) \leq q_1)$, which assigns weight 1 to the $\mathbf{z}_i$, for $i=1,...,n$,  with a robust squared Mahalanobis distance less than certain quantile $q_1$ of the chi-square distribution with $p+1$ degrees of freedom. The second weighting function is $wr_i=w(d^2 (r_i^{SW}))=I(d^2(r_i^{SW}) \leq q_2)$, which assigns weight 1 to the residuals $r_i^{SW}$ with a Mahalanobis distance less than certain quantile $q_2$ of the chi-square distribution with $1$ degree of freedom. 

The quantiles 
\begin{equation}\label{quanti}
q_1=\chi^2_{p+1,1-\delta_1} \quad \text{and}  \quad q_2=\chi^2_{1,1-\delta_2}
\end{equation}
\noindent
depend on the significance levels $\delta_1$ and $\delta_2$, for which $0.025$ and $0.01$ are chosen, respectively, as in \cite{rousseeuw2004robust}, because those are the classical choices for the threshold to detect outliers \citep{leroy1987robust}.

\section{Simulation structure}\label{sim}

In this section a simulation study is conducted to investigate the performance of the proposed SR regression estimator and compare it with OLS and some of the previously mentioned robust regression methods: LTS, MM, method S and REWLSE. The simulations were done in Matlab: OLS with the \textit{fitlm} function, LTS with the \textit{ltsregres} function from LIBRA library (see \cite{verboven2005libra}) considering the default option for the proportion of trimming which is $h=n/2 + 1$ and the default fraction of outliers the algorithm should resist which is equal to 0.75, MM with the \textit{MMreg} function from FSDA toolbox (see \cite{riani2012fsda}), with default values for the nominal efficiency: 0.95 and the rho function to weight the residuals as the bisquare which uses Tukey's functions,  method S with the function \textit{SEst} from the Discriminant Analysis Programme toolbox which computes biweight multivariate S-estimator for location and dispersion (see \cite{ruppert1992computing}) and REWLSE was computed with the functions the authors \cite{gervini2002class} kindly provided, with hard rejection weights and starting from an initial S-estimator.

Consider the linear regression model in matrix form:
\begin{equation}\label{model1}
\mathbf{y}=  \alpha +  X\boldsymbol{\beta}  + \boldsymbol{\epsilon}
\end{equation}

\noindent
where $X$ is of size $n\times p$, $\boldsymbol{\beta} = ( \beta_1, ...,\beta_p)^t $ is the unknown $p \times 1$ vector of regression parameters, $\alpha$ the unknown intercept, and the errors $ \boldsymbol{\epsilon}$ are i.i.d and independent from the carriers. The independent variables are distributed according to a multivariate standard Gaussian distribution $X \sim N(\mathbf{0}_{p},I_{p})$, where $\mathbf{0}_{p}$ is the $p-$dimensional vector of zeros and $I_p$ is the $p-$dimensional identity matrix. The simulation parameters are the following sets of dimension and sample size: $p=5$ with $n=20,30, 50, 100, 1000$, $p=20$ with $n=80, 100, 200, 500, 5000$ and $p=30$ with $n=100, 150, 300, 500, 5000$. The simulations are repeated $M=1000$ times and each time the parameter estimates are drawn anew. 

Three simulation scenarios are proposed, analogously as the simulation models found in the literature (\cite{maronna1986robust}, \cite{gervini2002class}, \cite{croux2003bounded},  \cite{rousseeuw2004robust}, \cite{agullo2008multivariate}, \cite{yu2017robust}).

\begin{itemize}
\item[(NE):] The response is generated from a standard normal distribution $N(0,I)$, which corresponds to putting $\boldsymbol{\beta}=\mathbf{0}$ and $\alpha=0$ when gaussian errors are considered.

\item[(TE):] The response is generated from a $t$-distribution with $3$ d.f, which corresponds to putting $\boldsymbol{\beta}=\mathbf{0}$ and $\alpha=0$ when $t_3$-distributed errors are considered.

\item[(NEO):] Normal errors as in [NE], but with probability $\delta$ the randomly selected observations in the independent variables were generated as $N(\lambda \sqrt{\chi^2_{p,0.99}} , 1.5)$ and the new response as $N(k\sqrt{\chi^2_{1,0.99}},1.5)$ where $\lambda,k= 0, 0.5 , 1 , 1.5, 2,3,4,5,6,7,8,9,10$.

\end{itemize}

For the last simulation scenario [NEO], the levels of contamination considered were $\delta=10\%, 20\%$.  Note that if $\lambda=0$ and $k>0$ we obtain \textit{vertical outliers}, if $\lambda>0$ and $k=0$ we obtain \textit{good leverage} points and if $\lambda>0$ and $k>0$ we obtain \textit{bad leverage} points. On the other hand, large values of $\lambda$ and $k$ produce extreme outliers, whereas small values produce intermediate outliers (see \cite{croux2003bounded} and \cite{agullo2008multivariate}).

\section{Efficiency}\label{seceff}

It is known that under simulation scheme [NE] the OLS estimator has maximum efficiency. The efficiency for each robust estimator, for finite samples, is calculated relative to OLS, considering the sum of squared deviations from the true coefficients and averaging over all repetitions. Consider the joint vector of regression parameters including the intercept $\boldsymbol{\varphi}=(\boldsymbol{\beta}^t,\alpha)^t$, which has dimension $(p+1)\times 1 $. For a certain robust method $R$, the finite sample efficiency for the joint estimator $\hat{\boldsymbol{\varphi}}_R$ is defined as:
\begin{equation}
\textrm{Eff}=\frac{1/M \sum_{m=1}^M || \hat{\boldsymbol{\varphi}}_{OLS}^{(m)} -\boldsymbol{\varphi}||_2^2}{1/M \sum_{m=1}^M || \hat{\boldsymbol{\varphi}}_R^{(m)} -\boldsymbol{\varphi}||_2^2}
\end{equation}

Table \ref{tab:effnormal} shows the simulated efficiencies relative to OLS, for the joint regression estimator $\hat{\boldsymbol{\varphi}}$ obtained with the proposed approach SR and the other robust regression methods, under simulation scheme [NE].

\begin{table}[H]
  \centering
  \caption{Finite sample efficiency in case of normal errors, scenario [NE]}
        \resizebox{10cm}{!}{ 
        \begin{tabular}{r|l|rrrrr}
    \toprule
    \multicolumn{1}{l|}{$p=5$} & $n$     & SR    & LTS   & S     & REWLSE & MM \\
    \midrule
          & 20    & \textbf{0.9182} & 0.2352 & 0.2715 & \textit{0.2346} & 0.2272 \\
          & 30    & \textbf{0.9828} & \textit{0.3486} & 0.4292 & 0.5026 & 0.4915 \\
          & 50    & \textbf{0.9833} & \textit{0.5061} & 0.5070 & 0.5129 & 0.5047 \\
          & 100   & \textbf{0.9839} & \textit{0.5870} & 0.7051 & 0.7441 & 0.7192 \\
          & 1000  & \textbf{0.9859} & \textit{0.7816} & 0.8691 & 0.9570 & 0.9159 \\
    \midrule
    \multicolumn{1}{l|}{$p=20$} & 80    & \textbf{0.9852} & 0.3763 & 0.6786 & \textit{0.2809} & 0.2963 \\
          & 100   & \textbf{0.9956} & \textit{0.3973} & 0.7966 & 0.5028 & 0.4955 \\
          & 200   & \textbf{0.9900} & \textit{0.4971} & 0.8630 & 0.5811 & 0.8015 \\
          & 500   & \textbf{0.9951} & \textit{0.6163} & 0.8719 & 0.8737 & 0.8393 \\
          & 5000  & \textbf{0.9981} & \textit{0.6822} & 0.9461 & 0.9611 & 0.9068 \\
             \midrule
    \multicolumn{1}{l|}{$p=30$} & 100   & \textbf{0.9900} & 0.4458 & 0.5068 & 0.3622 & \textit{0.2978} \\
          & 150   & \textbf{0.9927} & 0.4699 & 0.5155 & \textit{0.4347} & 0.5532 \\
          & 300   & \textbf{0.9933} & \textit{0.5110} & 0.5187 & 0.7524 & 0.5770 \\
          & 500   & \textbf{0.9970} & \textit{0.6467} & 0.8660 & 0.8479 & 0.8486 \\
          & 5000  & \textbf{0.9980} & \textit{0.6504} & 0.9646 & 0.9863 & 0.9781 \\
    \end{tabular}%
    }
  \label{tab:effnormal}%
\end{table}%

In each row, bold letter represent the higher efficiency and italic letter represent the lowest efficiency. 
The results show that for a fixed dimension, when the sample size is increased, all methods improve the resulting finite sample efficiency. LTS is the method that behaves poorly even when the sample size increases. S, REWLSE and MM require large samples in order to have efficiencies greater than $90\%$. The proposed method SR has higher efficiency for every dimension and sample sizes considered.

In the simulation scenario [TE], OLS is not a maximum efficient estimator, due to the heavy-tailed errors. Therefore, Table \ref{tab:tstudent} shows the mean squared errors (MSE) instead. The results show that, for all methods, large sample size translates into a decrease of the MSE, but method SR outperformed, in general, the other competitors.

\begin{table}[H]
  \centering
  \caption{MSE in case of $t-$student distributed errors, scenario [TE]}
    \resizebox{10cm}{!}{ 
    \begin{tabular}{r|l|rrrrr}
    \toprule
    \multicolumn{1}{l|}{$p=5$} & $n$     & \multicolumn{1}{l}{SR} & \multicolumn{1}{l}{LTS} & \multicolumn{1}{l}{S} & \multicolumn{1}{l}{REWLSE} & \multicolumn{1}{l}{MM} \\
    \midrule
          & 20    & \textbf{0.1499} & 0.2980 & 0.3634 & \textit{0.4892} & 0.3193 \\
          & 30    & \textbf{0.0579} & 0.0745 & 0.0662 & \textit{0.1074} & 0.0713 \\
          & 50    & \textbf{0.0304} & 0.0479 & 0.0409 & \textit{0.0548} & 0.0322 \\
          & 100   & \textbf{0.0114} & 0.0125 & \textit{0.0150} & 0.0115 & 0.0116 \\
          & 1000  & \textbf{0.0012} & 0.0016 & 0.0015 & \textit{0.0017} & 0.0014 \\
   \midrule
    \multicolumn{1}{l|}{$p=20$} & 80    & \textbf{0.0244} & 0.0443 & 0.0293 & \textit{0.1218} & 0.0881 \\
          & 100   & \textbf{0.0126} & 0.0376 & 0.0228 & \textit{0.0720} & 0.0364 \\
          & 200   & \textbf{0.0107} & 0.0108 & 0.0114 & 0.0117 & \textit{0.0118} \\
          & 500   & \textbf{0.0033} & \textit{0.0039} & 0.0036 & \textit{0.0039} & 0.0034 \\
          & 5000  & \textbf{0.0003} & \textit{0.0004} & \textit{0.0004} & \textbf{0.0003} & \textbf{0.0003} \\
    \midrule
    \multicolumn{1}{l|}{$p=30$} & 100   & \textbf{0.0202} & 0.0637 & 0.0375 & \textit{0.1767} & 0.0855 \\
          & 150   & \textbf{0.0110} & 0.0208 & 0.0157 & \textit{0.0328} & 0.0240 \\
          & 300   &\textbf{ 0.0052} & 0.0067 & 0.0074 & \textit{0.0075} & 0.0055 \\
          & 500   & \textbf{0.0032} & \textit{0.0040} & 0.0038 & 0.0039 & 0.0033 \\
          & 5000  & \textbf{0.0003} & \textit{0.0005} & \textit{0.0005} & \textbf{0.0003} & \textbf{0.0003} \\
    \end{tabular}%
    }
  \label{tab:tstudent}%
\end{table}%

\section{Robustness}\label{secrob}

\begin{figure}[H]
 \centering
   \includegraphics[width=0.9\textwidth]{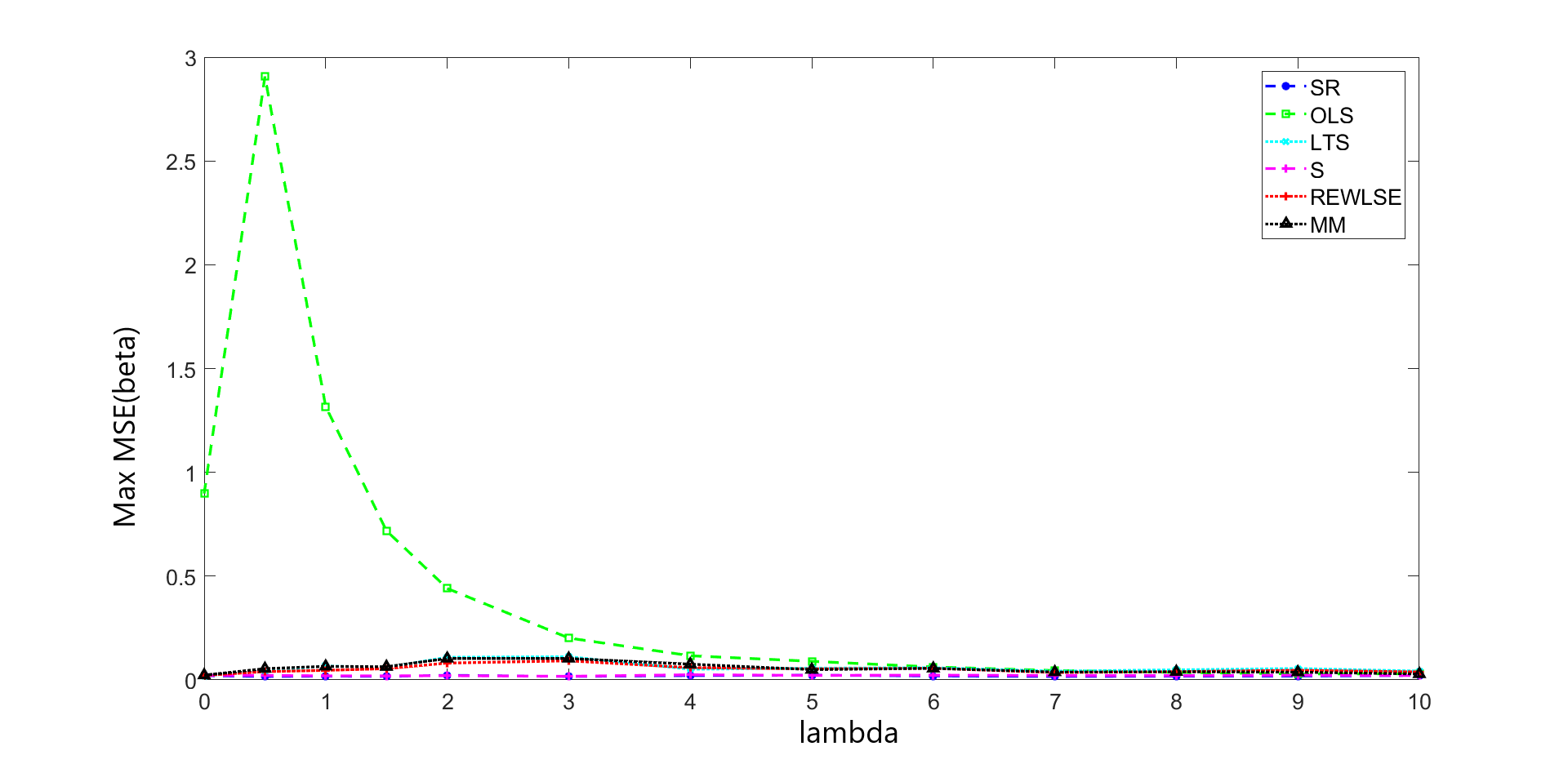}\\
   \caption{$MMSE(\hat{\boldsymbol{\beta}}$) with $p=5$, $n=100$, $\delta=10\%$.}
   \label{fig:msebeta_a10_p5}
\end{figure}

Simulations to study the robustness are carried out, considering the third simulation scheme [NEO]. The most significant results are those consisting on dimensions $p=5,30$ with sample sizes $n=100,500$, respectively. The two statistical criteria used to compare the estimators from the different approaches were the squared Bias and the MSE for the estimated parameter vector $\hat{\boldsymbol{\beta}}$ and for the estimated intercept $\hat{\alpha}$ averaging over all $M$ simulation runs (see \cite{gervini2002class}, \cite{croux2003bounded}, \cite{rousseeuw2004robust}). The following figures show, for each value of $\lambda$, the maximal value of MSE or Bias, obtained over all possible values of $k$. 
\begin{align}
MMSE_{\lambda} (\cdot) &= max_{k \in \{0,...,10\}} MSE_{\lambda,k} (\cdot)  \nonumber  \\
MBias_{\lambda} (\cdot) &= max_{k \in \{0,...,10\}} Bias_{\lambda,k} (\cdot)
\end{align}
\noindent
for each $\lambda\in \{0,...,10\}$. Figure \ref{fig:msebeta_a10_p5} shows the $MMSE(\hat{\boldsymbol{\beta}}$), in case of low dimension $p=5$ with sample size $n=100$ and when the data is contaminated with a level of $10\%$. OLS shows high MSE when the data contains atypical observations, specially for vertical outliers and bad leverage observations associated with the first values of $\lambda$. 

\begin{figure}[H]
 \centering
   \includegraphics[width=1\textwidth]{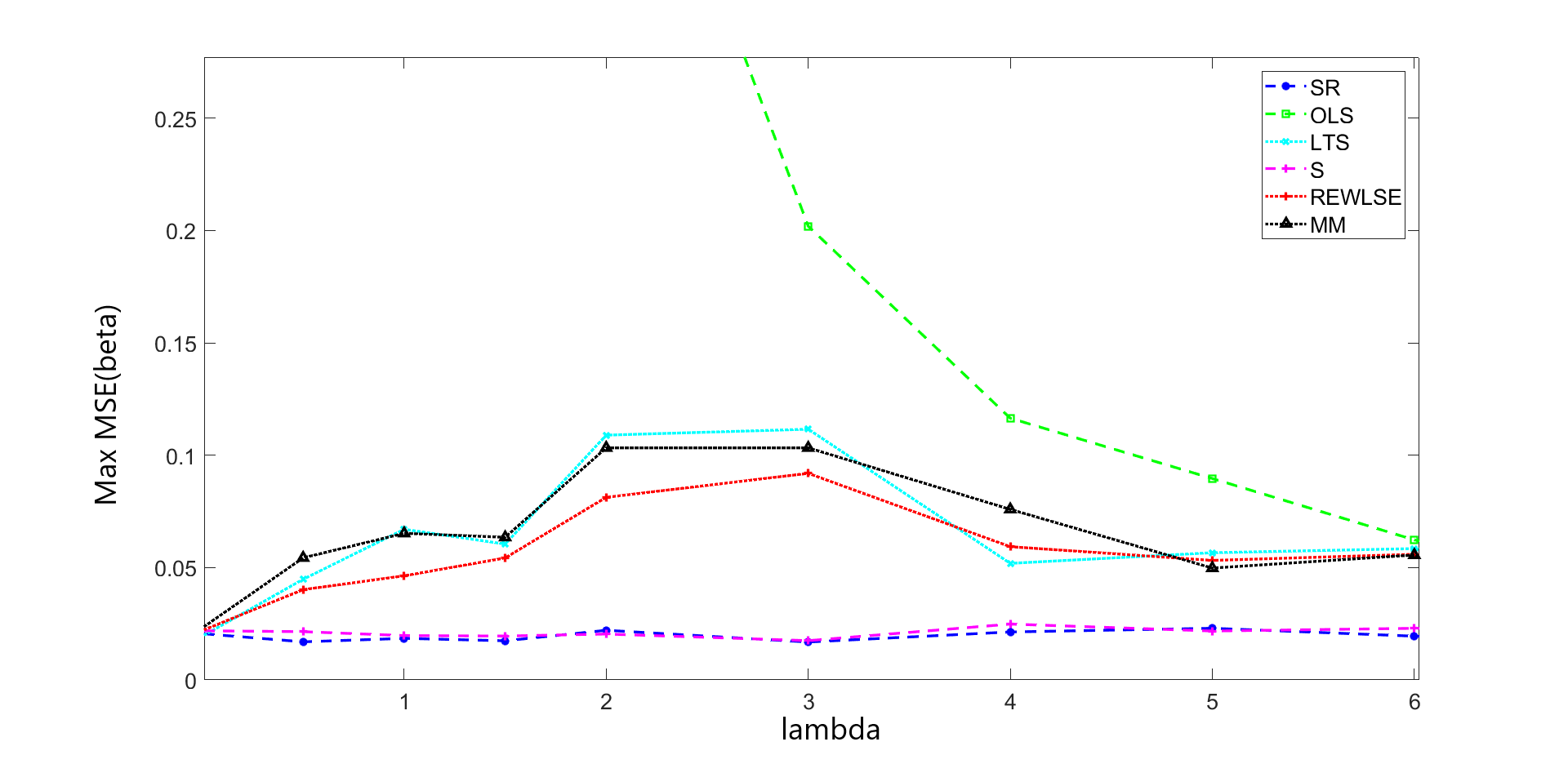}\\
   \caption{(Zoom) $MMSE(\hat{\boldsymbol{\beta}}$) with $p=5$, $n=100$, $\delta=10\%$.}
   \label{fig:msebeta_a10_p5zoom}
\end{figure}

If the previous image is zoomed, Figure \ref{fig:msebeta_a10_p5zoom}, it can be seen that for vertical outliers, i.e. $\lambda=0$, all robust methods have similar MSE, but for the remaining values of $\lambda$, the smallest errors correspond to the proposed method SR and method S. 

\begin{figure}[H]
 \centering
   \includegraphics[width=1\textwidth]{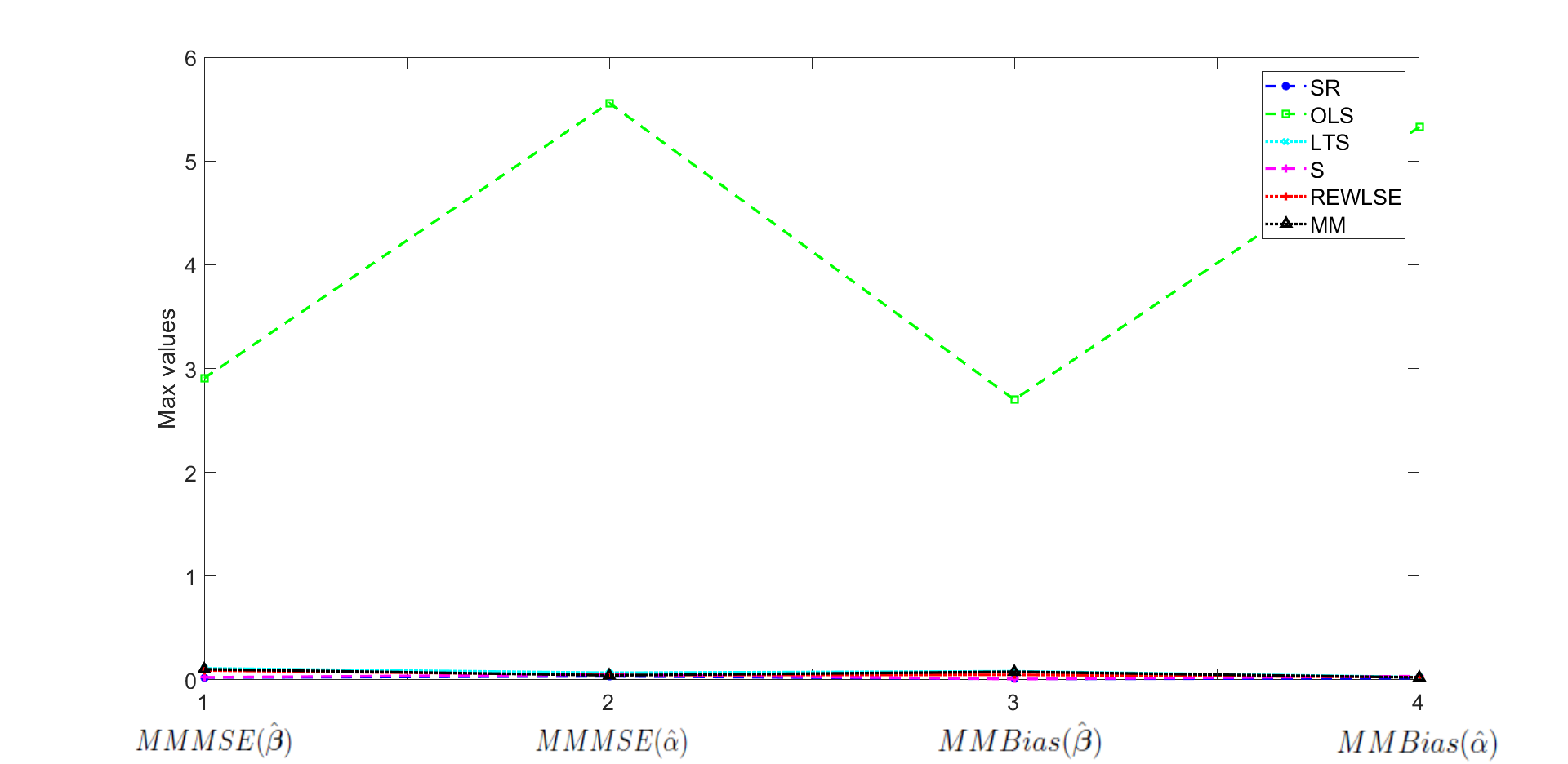}\\
   \caption{$MMMSE$,  with $p=5$, $n=100$ and  $\delta=10\%$.}
   \label{fig:alpha10_fig5_maxvalues1}
\end{figure}

For the MSE of $\hat{\alpha}$, and for the Bias of both $\hat{\alpha}$ and $\hat{\boldsymbol{\beta}}$, similar conclusions are obtained. In order to see these results from a different perspective, the error measures are summarized in a single graph for each dimension, sample size and contamination level. Figure \ref{fig:alpha10_fig5_maxvalues1} corresponds to $p=5$, $n=100$ and $\delta=10\%$. Each line represents a method. In the x-axis each number from 1 to 4 represents the maximum error measures: 1-MMMSE($\hat{\boldsymbol{\beta}}$), 2-MMMSE($\hat{\alpha}$), 3-MMBias($\hat{\boldsymbol{\beta}}$) and 4-MMBias($\hat{\alpha}$), over all possible values of $\lambda$.
\begin{align}
MMMSE (\cdot) &= max_{\lambda \in \{0,...,10\}} MMSE_{\lambda} (\cdot)  \nonumber  \\
MMBias (\cdot) &= max_{\lambda \in \{0,...,10\}} MBias_{\lambda} (\cdot)
\end{align}
\noindent
for each $\lambda\in \{0,...,10\}$.

\begin{figure}[H]
 \centering
   \includegraphics[width=1\textwidth]{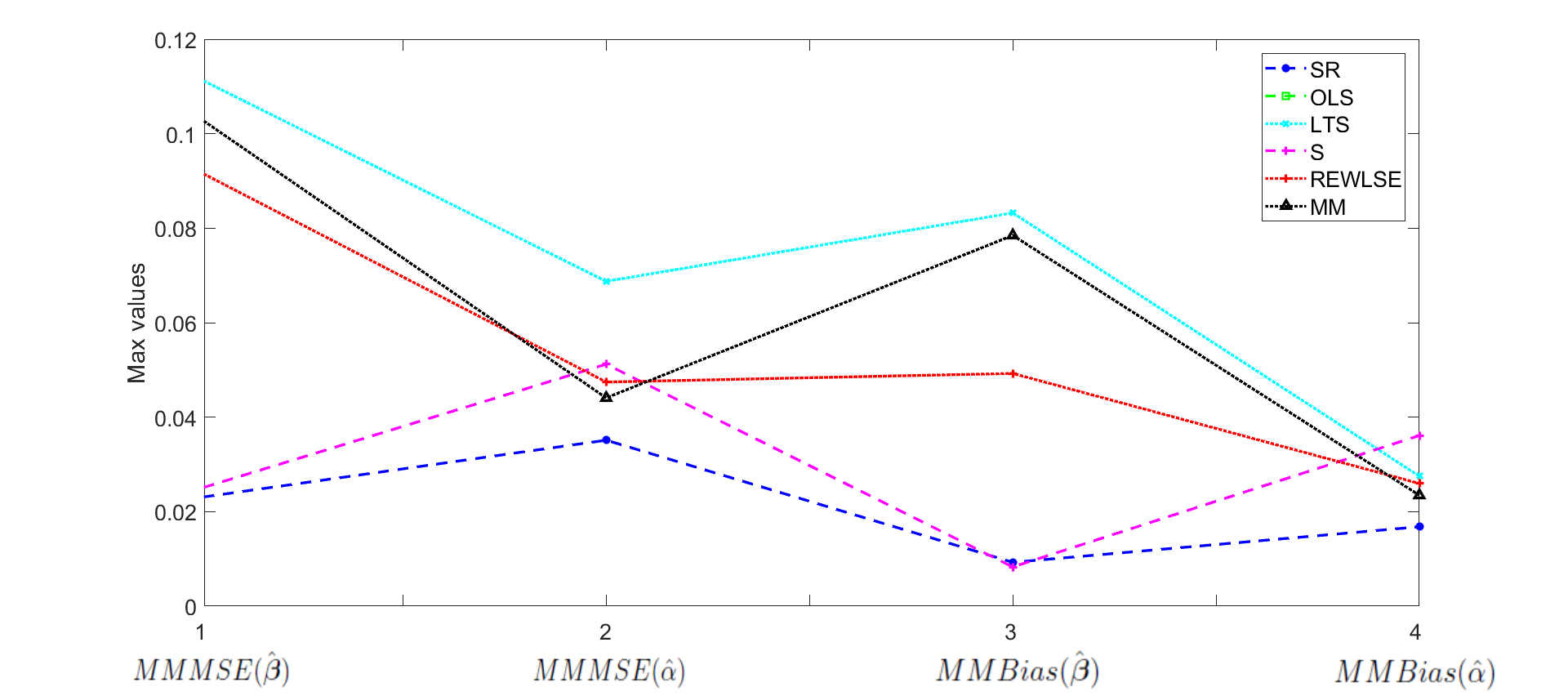}\\
   \caption{(Zoom) $MMMSE$, with $p=5$ and  $\delta=10\%$.}
   \label{fig:alpha10_fig6_maxvalues2}
\end{figure}

Figure \ref{fig:alpha10_fig6_maxvalues2} is a zoom of the previous Figure \ref{fig:alpha10_fig5_maxvalues1}. We can see in Figure \ref{fig:alpha10_fig6_maxvalues2} that in the majority of cases the proposed method SR has the lowest maximum MSE or Bias, except for one case in which method S has slightly lower maximum Bias($\hat{\boldsymbol{\beta}}$), but this happens only under low level of contamination.

\begin{figure}[H]
 \centering
   \includegraphics[width=1\textwidth]{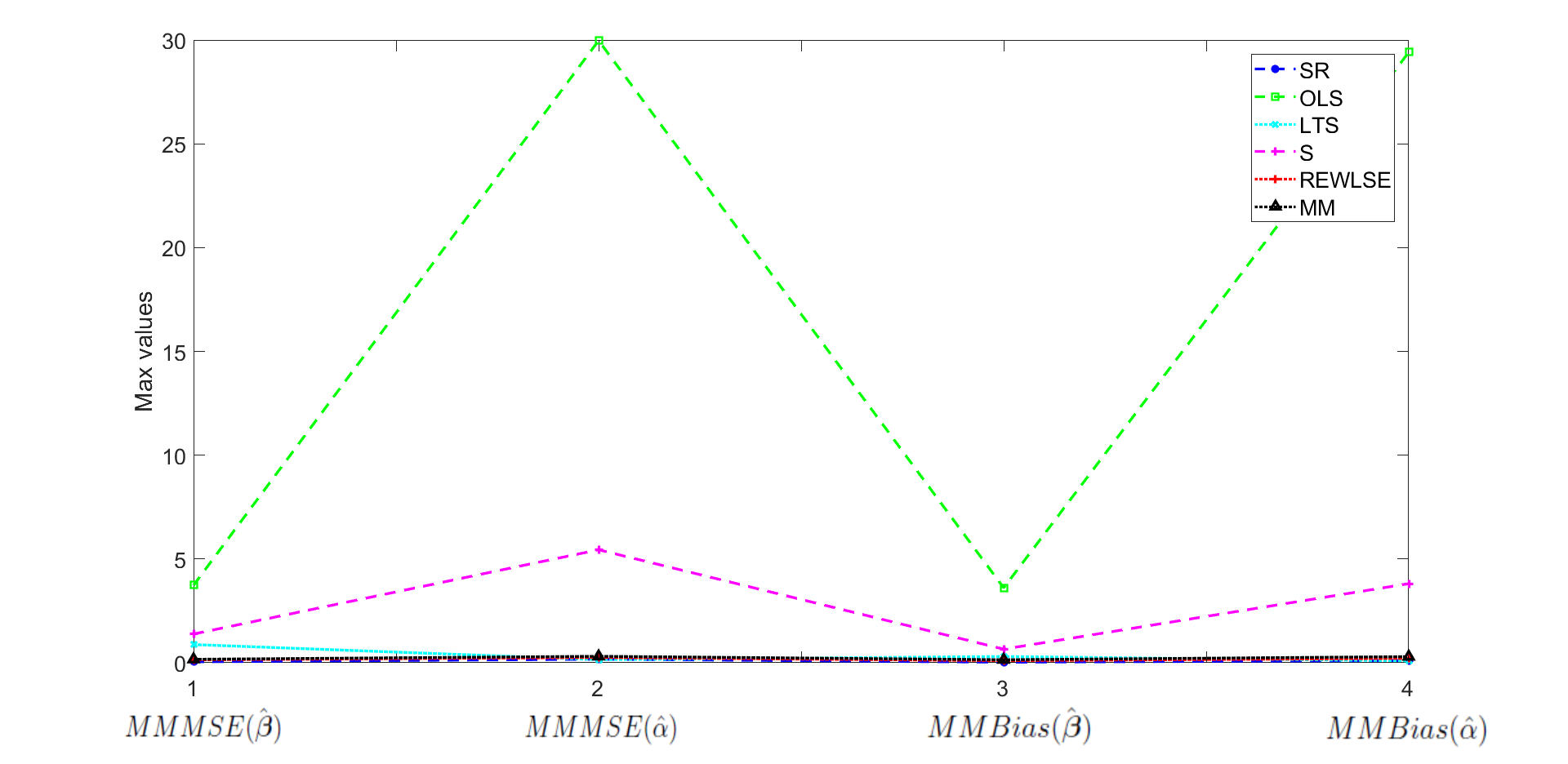}\\
   \caption{$MMMSE$, with $p=5$ and  $\delta=20\%$.}
   \label{fig:alpha20_fig5_maxvalues1}
\end{figure}

When the contamination level $\delta$ increases to $20\%$, method S worsens its performance as it can be seen in Figure \ref{fig:alpha20_fig5_maxvalues1}.
%

\begin{figure}[H]
 \centering
   \includegraphics[width=1\textwidth]{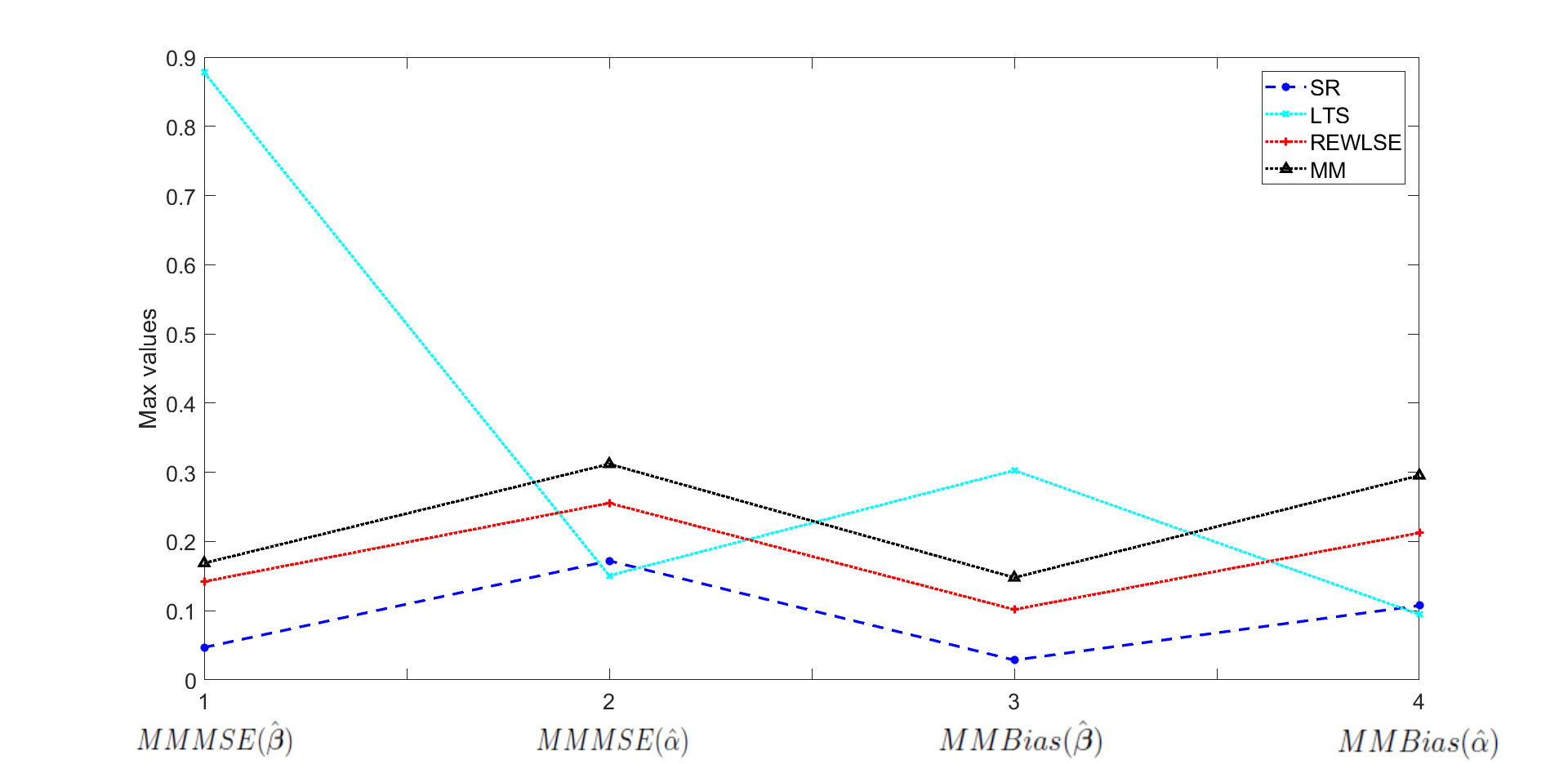}\\
   \caption{(Zoom) $MMMSE$, with $p=5$ and  $\delta=20\%$.}
   \label{fig:alpha20_fig6_maxvalues2}
\end{figure}

Zoomed Figure \ref{fig:alpha20_fig6_maxvalues2} shows that, in case of higher contamination level, SR is the overall best performance method taking into account that although MSE($\hat{\alpha}$) and Bias($\hat{\alpha}$) are slightly lower for LTS, the MSE and Bias of the $\hat{\boldsymbol{\beta}}$ for LTS is much higher than SR, REWLSE and even MM estimator.

%

\begin{figure}[H]
 \centering
   \includegraphics[width=1\textwidth]{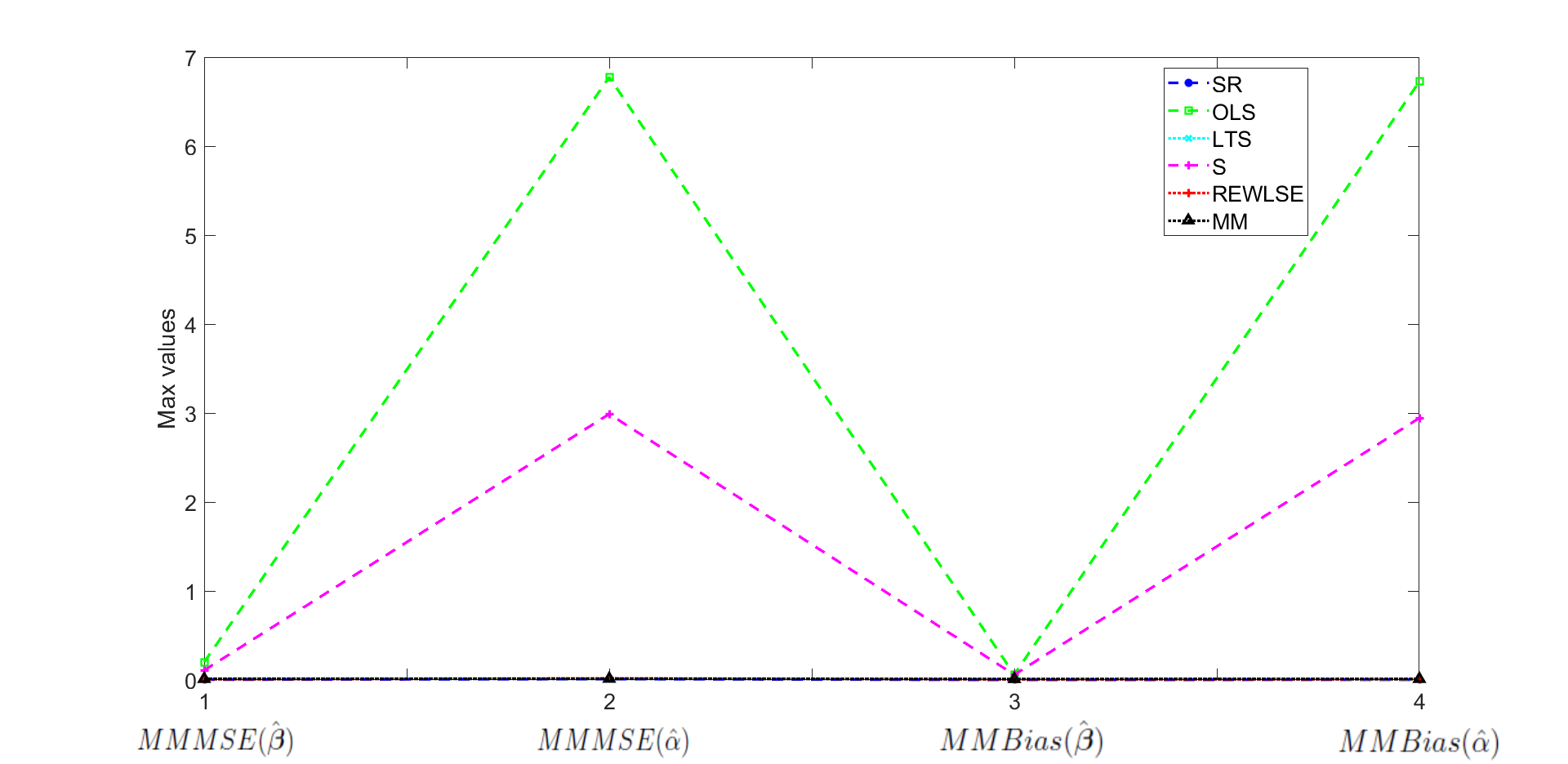}\\
   \caption{$MMMSE$, with $p=30$ and $\delta=10\%$. }
   \label{fig:30alpha10_fig5_maxvalues1}
\end{figure}

Figure \ref{fig:30alpha10_fig5_maxvalues1} shows that when the dimension is increased to $p=30$, and the contamination is $\delta=10\%$, the most affected methods are OLS and S. Method SR is the one that has the lowest maximum value for the MSE and Bias of both $\boldsymbol{\beta}$ and $\alpha$.

\begin{figure}[H]
 \centering
   \includegraphics[width=1\textwidth]{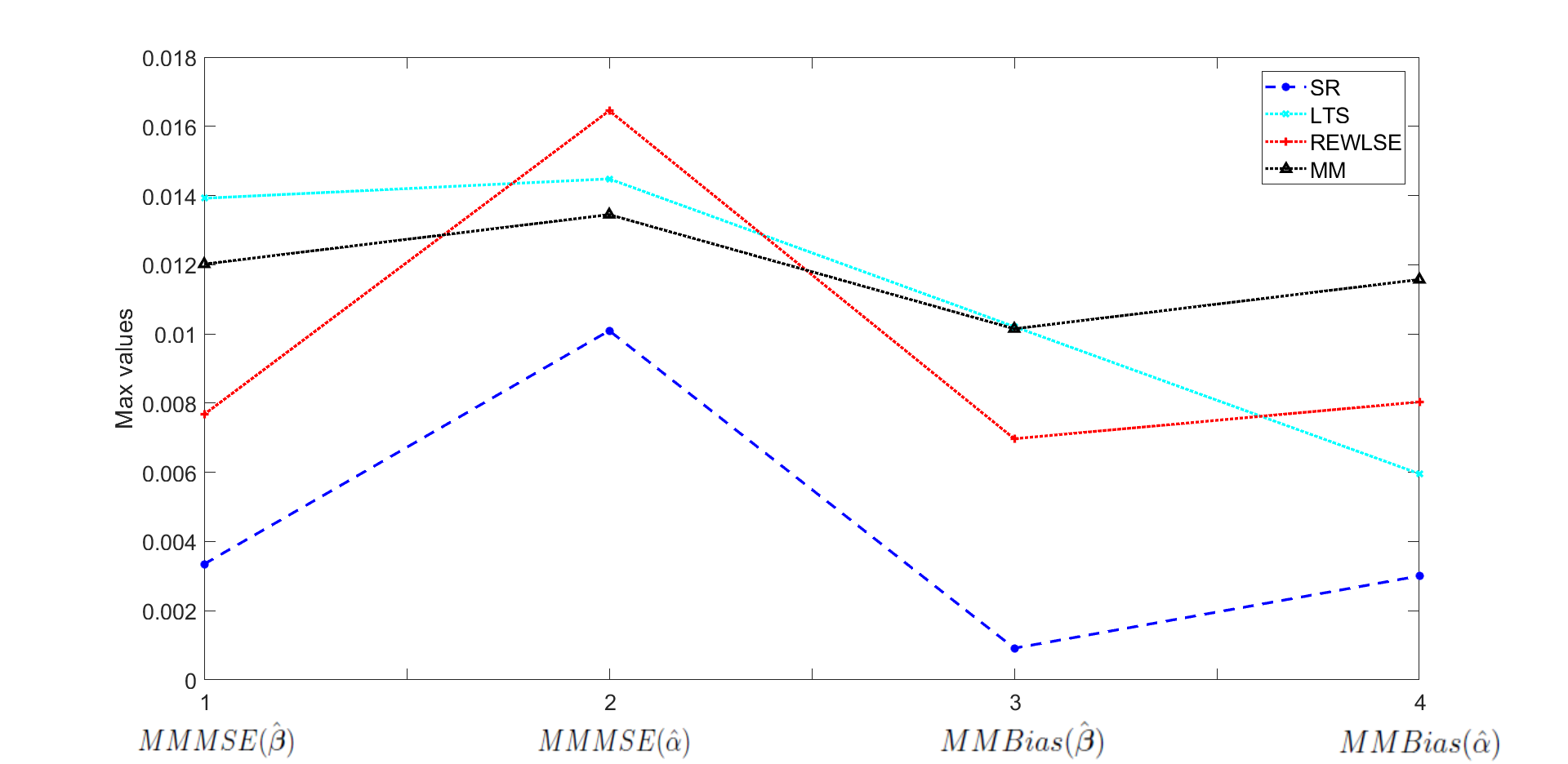}\\
   \caption{(Zoom) $MMMSE$, with $p=30$ and $\delta=10\%$.}
   \label{fig:30alpha10_fig6_maxvalues2}
\end{figure}

Figure \ref{fig:30alpha10_fig6_maxvalues2} is a zoom of Figure \ref{fig:30alpha10_fig5_maxvalues1} so we can see the four methods with lowest errors. A similar situation happens in case of $\delta=20\%$ of contamination. 

Tables \ref{tab:maxmaxp5a10} - \ref{tab:maxmaxp30a20} show the numerical results. For each method, the maximum (across $\lambda$ and $k$) MSE and Bias for both $\hat{\boldsymbol{\beta}}$ and $\hat{\alpha}$ for each combination of the dimension $p$ and the contamination level $\delta$, is showed. In bold letter are the lowest error and in italic letter are the highest error after OLS. The results bear out with the ones from the Figures.

%

\begin{table}[H]
  \centering
  \caption{MMMSE and MMBias of $\hat{\boldsymbol{\beta}}$ and $\hat{\alpha}$, for $p=5$ and $\delta=10\%$.}
     \resizebox{10cm}{!}{ 
     \begin{tabular}{lrrrr}
    \toprule
    Method & \multicolumn{1}{l}{MSE($\hat{\boldsymbol{\beta}}$)} & \multicolumn{1}{l}{MSE($\hat{\alpha}$)} & \multicolumn{1}{l}{BIAS($\hat{\boldsymbol{\beta}}$)} & \multicolumn{1}{l}{BIAS($\hat{\alpha}$)} \\
    \midrule
    OLS   & 2.9065 & 5.5593 & 2.7004 & 5.3280 \\
    SR    & \textbf{0.0230 }& \textbf{0.0351} & 0.0093 & \textbf{0.0168} \\   
    LTS   & \textit{0.1116} & \textit{0.0688} & \textit{0.0832} & 0.0275 \\
    S     & 0.0249 & 0.0512 & \textbf{0.0083} & \textit{0.0361} \\
    REWLSE & 0.0919 & 0.0474 & 0.0493 & 0.0260 \\
    MM    & 0.1033 & 0.0441 & 0.0785 & 0.0235 \\
    \end{tabular}%
    }
  \label{tab:maxmaxp5a10}%
\end{table}%

\begin{table}[H]
  \centering
  \caption{MMMSE and MMBias of $\hat{\boldsymbol{\beta}}$ and $\hat{\alpha}$, for $p=5$ and $\delta=20\%$.}
   \resizebox{10cm}{!}{ 
    \begin{tabular}{lrrrr}
    \toprule
    Method & \multicolumn{1}{l}{MSE($\hat{\boldsymbol{\beta}}$)} & \multicolumn{1}{l}{MSE($\hat{\alpha}$)} & \multicolumn{1}{l}{BIAS($\hat{\boldsymbol{\beta}}$)} & \multicolumn{1}{l}{BIAS($\hat{\alpha}$)} \\
    \midrule
     OLS   & 3.7360 & 29.9723 & 3.6101 & 29.4112 \\
    SR    & \textbf{0.0470} & 0.1720 & \textbf{0.0287} & 0.1075 \\
   LTS   & 0.8779 & \textbf{0.1508} & 0.3028 & \textbf{0.0947} \\
    S     & \textit{1.3853} & \textit{5.4441} & \textit{0.6577} & \textit{3.8112} \\
    REWLSE & 0.1422 & 0.2556 & 0.1018 & 0.2124 \\
    MM    & 0.1688 & 0.3120 & 0.1478 & 0.2954 \\
    \end{tabular}%
    }
  \label{tab:maxmaxp5a20}%
\end{table}%

\begin{table}[H]
  \centering
  \caption{MMMSE and MMBias of $\hat{\boldsymbol{\beta}}$ and $\hat{\alpha}$, for $p=30$ and $\delta=10\%$.}
    \resizebox{10cm}{!}{ 
    \begin{tabular}{lrrrr}
    \toprule
   Method & \multicolumn{1}{l}{MSE($\hat{\boldsymbol{\beta}}$)} & \multicolumn{1}{l}{MSE($\hat{\alpha}$)} & \multicolumn{1}{l}{BIAS($\hat{\boldsymbol{\beta}}$)} & \multicolumn{1}{l}{BIAS($\hat{\alpha}$)} \\
    \midrule
    OLS   & 0.1995 & 6.7748 & 0.0610 & 6.7250 \\
    SR    & \textbf{0.0033} & \textbf{0.0101} &\textbf{ 0.0009} & \textbf{0.0030} \\
    LTS   & 0.0139 & 0.0145 & 0.0102 & 0.0060 \\
    S     & \textit{0.1079} & \textit{2.9888} & \textit{0.0584} & \textit{2,9439} \\
    REWLSE & 0.0077 & 0.0165 & 0.0070 & 0.0080 \\
    MM    & 0.0120 & 0.0134 & 0.0101 & 0.0116 \\
    \end{tabular}%
    }
  \label{tab:maxmaxp30a10}%
\end{table}%

\begin{table}[H]
  \centering
  \caption{MMMSE and MMBias of $\hat{\boldsymbol{\beta}}$ and $\hat{\alpha}$, for $p=30$ and $\delta=20\%$.}
   \resizebox{10cm}{!}{ 
    \begin{tabular}{lrrrr}
    \toprule
    Method & \multicolumn{1}{l}{MSE($\hat{\boldsymbol{\beta}}$)} & \multicolumn{1}{l}{MSE($\hat{\alpha}$)} & \multicolumn{1}{l}{BIAS($\hat{\boldsymbol{\beta}}$)} & \multicolumn{1}{l}{BIAS($\hat{\alpha}$)} \\
    \midrule
    OLS   & 0.2317 & 25.5388 & 0.0639 & 25.3395\\
    SR    & \textbf{0.0044} & \textbf{0.0596} & \textbf{0.0011} & \textbf{0.0554} \\
    LTS   & 0.0450 & 0.3952 & 0.0400 & 0.3677 \\
    S     & \textit{0.1710} & \textit{15.0446} & \textit{0.0635} & \textit{14.8378} \\
    REWLSE & 0.0120 & 0.0980 & 0.0017 & 0.0930 \\
    MM    & 0.0356 & 0.1994 & 0.0262 & 0.1860 \\
    \end{tabular}%
    }
  \label{tab:maxmaxp30a20}%
\end{table}%

\subsection{Computational times}

The computational times in seconds for each method in simulation scenario [NEO] are also measured. The study was performed in a PC with a 3.40 GHz Intel Core i7 processor with 32GB RAM. The results are averaged for $10\%$ and $20\%$ of contamination since they were similar. OLS is obviously the fastest one because its simplicity. Following OLS, the proposed method SR is the second fastest method because it does not relies on iterative algorithms to calculate the estimations. The other robust competitors are between 3 and 9 times slower than our proposal SR for low dimension, and between 3 and 12 times slower for higher dimension.

\begin{table}[h]
  \centering
  \caption{Computational times with Normal distribution $p=5$ and $n=100$}
   \resizebox{10cm}{!}{ 
    \begin{tabular}{l|rrrrrr}
    $\alpha$ & SR & OLS &LTS &S& REWLSE & MM \\
    \midrule
    0.1   & 0.0206 & 0.0126 & 0.0989 & 0.0515 & 0.0572 & 0.1816 \\
    0.2   & 0.0200 & 0.0102 & 0.0966 & 0.0514 & 0.0545 & 0.1862 \\
    \end{tabular}%
    }
  \label{tab:compu1}%
\end{table}%

\begin{table}[htbp]
  \centering
  \caption{Computational times  with Normal distribution $p=30$ and $n=500$}
   \resizebox{10cm}{!}{ 
    \begin{tabular}{l|rrrrrr}
     $\alpha$ & SR & OLS &LTS &S& REWLSE & MM \\
    \midrule
    0.1   & 0.1246 & 0.0120 & 0.4350 & 0.3825 & 0.3967 & 1.5263 \\
    0.2   & 0.1209 & 0.0104 & 0.4102 & 0.3820 & 0.4192 & 1.5456 \\
    \end{tabular}%
    }
  \label{tab:compu2}%
\end{table}%
%
%
%
%
%
%
%
%

%
%
%

\section{Equivariance properties}\label{seceq}

The initial shrinkage robust estimators $\hat{\boldsymbol{\mu}}_{Sh}$ and $\hat{\Sigma}_{Sh}$ are approximately affine equivariant (\cite{cabana2019multivariate}). This means that the equivariance property cannot be demonstrated analytically because only part of the property holds, but it can be studied by means of simulations (as in \cite{Maronna2002} and \cite{Sajesh2012}). Then, the distance defined in Equation \ref{eq:primeradist} and used in the weights for the SW estimators of mean and covariance matrix (Equation \ref{eq:locshrinkrewest2}) remains approximately invariant under affine transformations. Since the weights are hard rejection depending on the robust distance, the estimators $\hat{\boldsymbol{\mu}}_{Sh}^{SW}$ and $\hat{\Sigma}_{Sh}^{SW}$ should hold the property. However, the real interest in the regression problem is concerned around the parameter estimators, denoted as: 
$\hat{\boldsymbol{\varphi}}^{SR} = \left ( \left (\hat{\boldsymbol{\beta}}^{SR}\right )^t, \hat{\alpha}^{SR}   \right )^t$. Thus, we propose to study the equivariance property on them. Affine equivariance in regression can be split in the three following properties (\cite{rousseeuw2004robust} and \cite{maronna1986robust}):

\begin{enumerate}
\item \textbf{Regression equivariance:}
If a linear function of the explanatory variables is added to the response, then the coefficients of this linear function are also added to the estimators.

\item \textbf{y-equivariance:}
If the response variable is transformed linearly then the estimators transforms correctly.

\noindent
Property (1) and (2) can be seen together as:

\begin{equation}\label{yequiv}
\hat{\boldsymbol{\varphi}}^{SR}(X, \mathbf{y}c+X\mathbf{g}+v)=\hat{\boldsymbol{\varphi}}^{SR}(X, \mathbf{y})c + (\mathbf{g}^t,v)^t
\end{equation}
\noindent
where $c \in \mathbb{R}$ is any non-singular constant, $\mathbf{g}$ is any $p \times 1 $ vector and $v \in \mathbb{R}$ is any constant.
This means that, keeping the same $X$, and transforming the response as $\mathbf{y}c+X\mathbf{g}+v$, the resulting transformed estimators are:  $\hat{\boldsymbol{\beta}}^{SR}_{new}=c(\hat{\boldsymbol{\beta}}^{SR})+\mathbf{g}$ and $\hat{\alpha}^{SR}_{new}= c\hat{\alpha}^{SR}+v$.

\item[3.] \textbf{x-equivariance:}
Also called \textit{carrier equivariance}. It says that if the explanatory variables are transformed linearly (coordinate system transformation), then the estimators transforms correctly.

\begin{equation}\label{xequiv}
\hat{\boldsymbol{\varphi}}^{SR}(XA, \mathbf{y})=((\hat{\boldsymbol{\beta}}^{SR})^t (A^{-1})^t,\hat{\alpha}^{SR})^t
\end{equation}

This means that if the carriers are transformed as $XA$ with any non-singular $p\times p$ matrix $A$, the resulting estimators are: $\hat{\boldsymbol{\beta}}^{SR}_{new}=A^{-1}\hat{\boldsymbol{\beta}}^{SR}$ and the intercept should remain the same $\hat{\alpha}^{SR}_{new}=\hat{\alpha}^{SR}$. 
\end{enumerate}

Exploring all possible transformations is infeasible, that is the reason why \cite{Maronna2002} and \cite{Sajesh2012} proposed to generate the random matrices $A$ for the \textbf{x-equivariance} as $A = TD$, where $T$ is a random orthogonal matrix and $D = diag(u_1,...,u_p)$, where the $u_j$'s are independent and uniformly distributed in $(0,1)$, for all $j=1,...,p$. Then, each generated data matrix $X$ in each repetition, is transformed with a random transformation $A$. Following this idea, we propose to generate the non-singular $c$, the $\mathbf{g}$ and the $v$ for regression and \textbf{y-equivariance}, randomly for each repetition.

The MSE of the proposed method SR is studied when the transformations described above are made to the simulated data-set. Consider the simulation scenario [NE] for normal data without outliers ($\delta=0\%$) and scenario [NEO] when there is $\delta=10\%,20\%$ of contamination, to see the impact of the presence of outliers. The vector of regression parameters $\hat{\boldsymbol{\varphi}}^{SR}$ is estimated with the untransformed data and saved. After that, the data is transformed according to Equation \ref{yequiv} for the regression and $\mathbf{y}$-equivariance and according to Equation \ref{xequiv} for the $\mathbf{x}$-equivariance. Next, the method SR is applied to the transformed data and the resulting $\hat{\boldsymbol{\varphi}}^{SR}_{new}$ are saved. The MSE is calculated between the obtained $\hat{\boldsymbol{\varphi}}^{SR}_{new}$ and what it should be obtained if the equivariance properties hold. Table \ref{tab:yequiv} shows for each $\lambda$, the resulting $MMSE_{\lambda}(\hat{\boldsymbol{\varphi}}^{SR}_{new})$.

\begin{table}[H]
  \centering
  \caption{$MMSE_{\lambda}(\hat{\boldsymbol{\varphi}}^{SR}_{new})$ for regression and $\mathbf{y}$-equivariance}
       \resizebox{10cm}{!}{
       \begin{tabular}{l|lll|lll}
          & \multicolumn{1}{l}{$p=5$} &     &    & \multicolumn{1}{l}{$p=30$} &  &  \\
    \midrule
    $\lambda$  & \multicolumn{1}{l}{$\delta=0\%$}  & \multicolumn{1}{l}{$\delta=10\%$} & \multicolumn{1}{l|}{$\delta=20\%$} &  \multicolumn{1}{l}{$\delta=0\%$} & \multicolumn{1}{l}{$\delta=10\%$} & \multicolumn{1}{l}{$\delta=20\%$} \\
    \midrule
    0     & 0.01205  & 0.04173 & 0.12625 & 0.00006 & 0.26366 & 0.30312 \\
    0.5     &0.00567 & 0.01994 & 0.03135 & 0.00009 & 0.00267 & 0.00085 \\
    1       & 0.00645 & 0.01206 & 0.00876 & 0.00005 & 0.00272 & 0.00066 \\
    1.5    & 0.00615 & 0.00924 & 0.00373 & 0.00009 & 0.00428 & 0.00046 \\
    2       &0.00686 & 0.00822 & 0.00384 & 0.00008 & 0.00156 & 0.00037 \\
    3       &0.01718 & 0.00521 & 0.00454 & 0.00008 & 0.00215 & 0.00057 \\
    4       & 0.00726& 0.00905 & 0.00756 & 0.00008 & 0.00298 & 0.00068 \\
    5       & 0.00863 & 0.01228 & 0.00737 & 0.00007 & 0.00208 & 0.00063 \\
    6       &0.00586 & 0.01305 & 0.00677 & 0.00004 & 0.00166 & 0.00034 \\
    7      & 0.00822 & 0.00934 & 0.00550 & 0.00003 & 0.00265 & 0.00044 \\
    8      &0.00707  & 0.01955 & 0.00628 & 0.00007 & 0.00227 & 0.00056 \\
    9      & 0.00545 & 0.00948 & 0.01328 & 0.00002 & 0.00306 & 0.00077 \\
    10     & 0.00676 & 0.02298 & 0.00686 & 0.00009 & 0.00409 & 0.00037 \\
    \end{tabular}%
    }
  \label{tab:yequiv}%
\end{table}%

For vertical outliers, i.e. when $\lambda=0$, the error increases with the increase in dimension and contamination level, a fact that is influenced mostly by the error of the intercept. Nevertheless, for the rest of the cases the maximum possible error is low. Table \ref{tab:xequiv} shows the results for the $\mathbf{x}$-equivariance. In this case, both for vertical outliers and leverage points, the error remains low. Thus, since the errors are mostly controlled, the proposed robust regression estimator is approximately regression, y- and x-equivariant.

\begin{table}[H]
  \centering
  \caption{$MMSE_{\lambda}(\hat{\boldsymbol{\varphi}}^{SR}_{new})$ for $\mathbf{x}$-equivariance}
     \resizebox{10cm}{!}{
       \begin{tabular}{l|lll|lll}
          & \multicolumn{1}{l}{$p=5$} &  &    & \multicolumn{1}{l}{$p=30$} & & \\
    \midrule
    $\lambda$  & \multicolumn{1}{l}{$\delta=0\%$}  & \multicolumn{1}{l}{$\delta=10\%$} & \multicolumn{1}{l|}{$\delta=20\%$} &  \multicolumn{1}{l}{$\delta=0\%$} & \multicolumn{1}{l}{$\delta=10\%$} & \multicolumn{1}{l}{$\delta=20\%$} \\
    \midrule
    0     & 0.00206  & 0.00421 & 0.01874 & 0.00005  & 0.01324 & 0.09468 \\
    0.5   & 0.00162 & 0.00456 & 0.01310 & 0.00003  & 0.00026 & 0.00008 \\
    1      &0.00178 & 0.00348 & 0.00493 & 0.00003 & 0.00030 & 0.00003 \\
    1.5    &0.00153 & 0.00392 & 0.00132 & 0.00004 & 0.00012 & 0.00006 \\
    2      &0.00198 & 0.00320 & 0.00234 & 0.00003 & 0.00034 & 0.00003 \\
    3      &0.00144 & 0.00293 & 0.00208 & 0.00003 & 0.00016 & 0.00002 \\
    4      &0.00177 & 0.00329 & 0.00359 & 0.00005 & 0.00026 & 0.00005 \\
    5      &0.00194 & 0.00339 & 0.00182 & 0.00003 & 0.00020 & 0.00001 \\
    6     & 0.00173 & 0.00481 & 0.00205   & 0.00005 & 0.00016 & 0.00002 \\
    7     &0.00214  & 0.00329 & 0.00184 & 0.00002 & 0.00012 & 0.00002 \\
    8     & 0.00186 & 0.00415 & 0.00177 & 0.00004 & 0.00013 & 0.00002 \\
    9      & 0.00242& 0.00356 & 0.00188 & 0.00004 & 0.00016 & 0.00001 \\
    10    & 0.00193 & 0.00287 & 0.00250 & 0.00003 & 0.00011 & 0.00001 \\
    \end{tabular}%
    }
  \label{tab:xequiv}%
\end{table}%

\section{Breakdown property}\label{secbdp}

The bdp measures the maximum proportion of outliers that the estimator can safely tolerate. The highest possible value for the bdp is $50\%$. The empirical breakdown value can be examined through simulations, as in \cite{Sajesh2012}, considering high contamination levels. Although these situations are not that relevant in practice because low levels of contamination should be expected, we propose to study if the error and the bias are controlled in these scenarios in order to see the performance of the proposed SR estimator. For this, [NEO] contamination scheme is used, but considering higher levels of contaminations $\delta=30\%, 40\%, 45\%$. Table \ref{tab:bdp5} shows the resulting MMMSE and MMBias for $\hat{\boldsymbol{\varphi}}^{SR}_{new}$ in the low dimension $p=5$ case.

\begin{table}[H]
  \centering
  \caption{MMMSE and MMBias, $p=5$}
     \resizebox{11cm}{!}{
     \begin{tabular}{l|ll|ll|ll}
        & $\delta=30\%$  &       & $\delta=40\%$  &       & $\delta=45\%$   &  \\
    \midrule
    Method & MMMSE   & MMBias  & MMMSE   & MMBias  & MMMSE   & MMBias  \\
    \midrule
    OLS   & 6.9013 & 5.9143 & 7.5851 & 6.3344 & 7.6727 & 6.3215 \\
    SR    & \textbf{0.1216} & \textbf{0.1160} & \textbf{0.2733} & \textbf{0.1343} & \textbf{0.3314} & \textbf{0.1301} \\
    LTS   & 6.0032 & 5.6686 & 6.6864 & 6.3431 & 6.9428 & 6.4081 \\
    S     & \textit{6.0679} & \textit{5.7893} & \textit{7.2842} & \textit{7.0237} & \textit{7.2403} & \textit{6.7814} \\
    REWLSE & 0.3251 & 0.2994 & 1.0797 & 0.7422 & 1.7883 & 1.0121 \\
    MM    & 0.5190 & 0.4884 & 1.4912 & 1.1475 & 3.6681 & 2.6982 \\
    \end{tabular}%
    }
  \label{tab:bdp5}%
\end{table}%

\begin{table}[H]
  \centering
  \caption{MMMSE and MMBias, $p=30$}
       \resizebox{11cm}{!}{
       \begin{tabular}{l|ll|ll|ll}
        & $\delta=30\%$  &       & $\delta=40\%$  &       & $\delta=45\%$   &  \\
    \midrule
    Method & MMMSE   & MMBias  & MMMSE   & MMBias  & MMMSE   & MMBias  \\
    \midrule
    OLS   & 1.2970 & 1.0677 & 1.3839 & 1.0666 & 1.2738 & 1.0701 \\
    SR    & \textbf{0.0131} & \textbf{0.0025} & \textbf{0.0642} & \textbf{0.0182} & \textbf{0.1138} & \textbf{0.0232} \\
    LTS   & \textit{0.6640} & \textit{0.1567} & \textit{1.0824} & \textit{0.2211} & \textit{0.9589} & \textit{0.1980} \\
    S     & 0.2660 & 0.0678 & 0.3764 & 0.0665 & 0.3042 & 0.0749 \\
    REWLSE & 0.0218 & 0.0034 & 0.0977 & 0.0310 & 0.2184 & 0.0630 \\
    MM    & 0.0732 & 0.0677 & 0.2274 & 0.0668 & 0.4012 & 0.0675 \\
    \end{tabular}%
    }
  \label{tab:bdp30}%
\end{table}%

Table \ref{tab:bdp30} shows the results for higher dimension $p=30$. Bold letter represents lower error or bias and italic letter represents the highest measures after OLS, which is the method with worse results. LTS, S and MM have high error and bias for both low and high dimension, specially with the increase of the contamination level. REWLSE is competitive with SR in high dimension, but in low dimension REWLSE shows higher errors. The MSE and Bias of SR remain low, specially in high dimension and even with large contamination in the data, compared with the other robust methods supposedly having a high bdp. As discussed in \cite{yu2017robust} where the authors review and compare some robust regression approaches, the issue here is that although LTS, S and MM have high bdp, the computation is very challenging (\cite{hawkins2002inconsistency} and \cite{stromberg2000least}). That is why resampling algorithms are used to obtain a number of subsets and then compute the robust regression estimate from a number of initial estimates. However, the high breakdown property usually requires that the number of elementary sets goes to infinity, for example, \cite{hawkins2002inconsistency} proved that LTS computed with fast-LTS algorithm had zero bdp. In order to compute these  estimators with high bdp, one should consider all possible elemental sets. SR approach shows high resistance to large contamination even in high dimension, which can be translated in high empirical bdp.

\section{Real examples}\label{realex}

In this section, we study two known data-sets, very often used in the literature, to illustrate the performance of the proposed robust regression method comparing to the other robust alternatives. And also a socioeconomic and environmental related data-set that explains the Living Environment Deprivation of areas of Liverpool through remote-sensed data obtained from Google Earth technologies (\cite{arribas2017remote}).

\subsection{Star data}

\begin{figure}[H]
 \centering
   \includegraphics[width=1\textwidth]{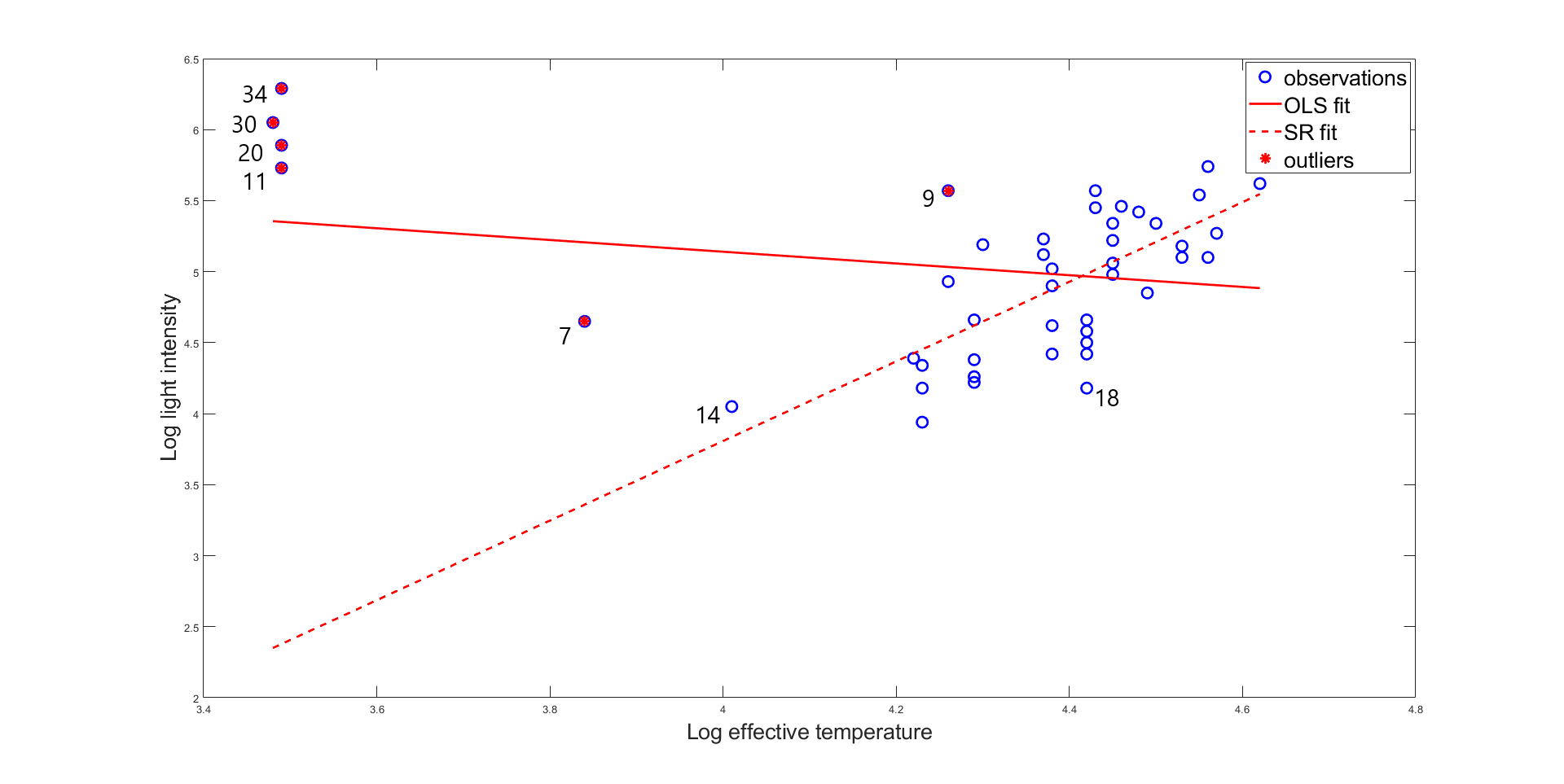}\\
   \caption{Star data-set with OLS and SR regression fit.}
   \label{fig:starsolsSR}
\end{figure}

The first example is the star data-set, and it is reported in \cite{leroy1987robust}, and based on \cite{humphreys1978studies} and  \cite{vansina1982close}. It has become a bench-mark for robust regression methodologies. It consists on $n=47$ observations corresponding to 47 stars of the CYG OB1 cluster in the direction of Cygnus. There is only one carrier $x$ which is the logarithm of the effective temperature at the surface of the star. The response variable $y$ is the logarithm of its light intensity. There is a positive linear relationship between the response and the explanatory variables, except for four red giant stars (observations 11, 20, 30 and 34) which are outliers because they have low temperatures and a high output of light (the four observations on the upper left corner in Figure \ref{fig:starsolsSR}). 

These giant stars actually represent a different population. They are bad leverage points because they influence OLS regression line due to the poor estimation of the parameters. Figure \ref{fig:starsolsSR} shows how the four giant stars pull the OLS line towards them. Observations 7 and 9 are intermediate outliers. And finally, in the multivariate sense, observation 14 is often detected as outlier, but in the regression sense it is a good leverage point because it follows the same linear pattern than the bulk data. Robust regression fit made by the proposed method SR detected the giant stars 11, 20, 30, 34 and the intermediate outliers 7 and 9.

Table \ref{tab:starestimates} summarizes all method's estimation of the intercept and slope, and the outliers detected by the robust methods. Note that OLS estimates are completely changed, they have even different sign. SR and REWLSE correctly detect the regression outliers, method S detects the good leverage point, observation 14, as an outlier. LTS detects observation 18 as atypical when it is not. In Figure \ref{fig:starsolsSR} it can be seen that observation 18 is an example of the swamping effect problem. On the other hand, MM approach only detects as outliers the giant stars (masking effect).

\begin{table}[H]
  \centering
  \caption{Estimation of intercept and slope and detected outliers with star data.}
         \resizebox{7.5cm}{!}{
         \begin{tabular}{l|rr|l}
    Method & $\hat{\alpha}$ & $\hat{\beta}$ & Detected outliers    \\
    \midrule
    OLS & 6.7935 & -0.4133  &    \\
    SR  & -7.4035 & 2.9028  &  7  9  11  20  30  34 \\
    LTS & -8.5001 & 3.0462 & 7  9  11  18  20  30  34 \\
    S  & -10.5034 & 3.4994 & 7  9  11  14  20  30  34 \\
    REWLSE & -7.5001 & 3.0462  & 7  9  11  20  30  34 \\
    MM  & -5.1234 & 2.2879 & 11  20  30  34 \\
    \end{tabular}%
    }
  \label{tab:starestimates}%
\end{table}%

The $R^2$ values for the linear regression models fitted by each method are summarized in Table \ref{tab:r2stars}. OLS's coefficient of determination is low, while that of the robust methods is high, except for MM approach which is lower than the rest.

\begin{table}[H]
  \centering
  \caption{$R^2$ for each method with stars data-set.}
     \resizebox{10cm}{!}{
     \begin{tabular}{c|c|c|c|c|c|c}
          Method & OLS & SR & LTS & S & REWLSE & MM   \\
    \midrule
    $R^2$    &  0.0443 & 0.7113 & 0.7006 & 0.7035 &0.7095 & 0.5578 \\
    \end{tabular}%
    }
  \label{tab:r2stars}%
\end{table}%

\subsection{Hawkins-Bradu-Kass data}

HBK data-set was artificially created by \cite{hawkins1984location} and it was also used in \cite{leroy1987robust}, and many others. It contains $p=3$ explanatory variables and a response variable. The first 14 observations are leverage points: 1-10 of bad type and 11-14 of good type. Thus, only observations 1-10 are outliers in the regression sense. Table \ref{tab:hbkestimates} shows the estimation by all methods for the three parameters, and it can be seen that OLS is highly influenced by the presence of these leverage points. Also, the parameter estimated by S method are different than that of the other robust approaches, and the reason for this is that all robust methods correctly detect the true outliers, except for method S, which also includes the good leverage points 11-14.

\begin{table}[H]
  \centering
  \caption{Estimation of the parameters and detected outliers with HBK data.}
    \resizebox{13cm}{!}{
     \begin{tabular}{l|cccc|l}
    Method          & $\hat{\beta}_0$ & $\hat{\beta}_1$ & $\hat{\beta}_2$ & $\hat{\beta}_3$ & Detected outliers    \\
    \midrule
    OLS      & -0.3875 & 0.2392 & -0.3345 & 0.3833 &    \\
    SR    & -0.1800     & 0.0836 & 0.0396 & -0.0518  & 1 2 3 4 5 6 7 8 9 10 \\
    LTS      & -0.1805     & 0.0814 & 0.0399 & -0.0517  & 1 2 3 4 5 6 7 8 9 10 \\
    S   & -0.0174  & 0.0957 & 0.0041 & -0.1286 & 1 2 3 4 5 6 7 8 9 10 11 12 13 14 \\
    REWLSE & -0.1805 & 0.0814 & 0.0399 & -0.0517  & 1 2 3 4 5 6 7 8 9 10 \\
    MM   & -0.1913 & 0.0860 & 0.0412 & -0.0541  & 1 2 3 4 5 6 7 8 9 10 \\
    \end{tabular}%
    }
  \label{tab:hbkestimates}%
\end{table}%

%
%
%
%

The adjusted $R^2$ values are summarized in Table \ref{tab:r2hbk}. Here, all robust methods, except S, have high and similar $R^2$.

\begin{table}[H]
  \centering
  \caption{Adjusted $R^2$ for each method with HBK data-set.}
     \resizebox{10cm}{!}{
     \begin{tabular}{c|c|c|c|c|c|c}
          Method & OLS & SR & LTS & S & REWLSE & MM   \\
    \midrule
    $R^2$    &  0.5850 & 0.9818 & 0.9816 & 0.9002 &0.9817 & 0.9811 \\
    \end{tabular}%
    }
  \label{tab:r2hbk}%
\end{table}%

\subsection{Living Environment Deprivation data}

In \cite{arribas2017remote}, the authors studied the Living Environment Deprivation (LED) index. This measure allows to study quantitatively the concept of quality of the local environment, known also as urban quality of life, which is a qualitative concept. This is an essential matter for environmental research, citizens and politics. This kind of indices can be explained through remote sensing data, i.e. information collected without making physical contact, for example, from satellite technologies. The authors in \cite{arribas2017remote} proposed to model the LED index of Liverpool (UK) based on four sets of explanatory variables extracted from a very high spatial resolution (VHR) image downloaded from Google Earth. The four groups are called: land cover (LC), spectral (SP), texture (TX) and structure features (ST). See \cite{arribas2017remote} for more detailed description of the features. The authors first propose to explain the LED index with a linear combination of the four sets of variables. The linear regression model is the following:
\begin{equation}
LED=\alpha + \boldsymbol{\beta} LC + \boldsymbol{\gamma} SP + \boldsymbol{\delta} TX + \boldsymbol{\zeta} ST + \epsilon
\end{equation}

\begin{figure}[H]
 \centering
   \includegraphics[width=0.8\textwidth]{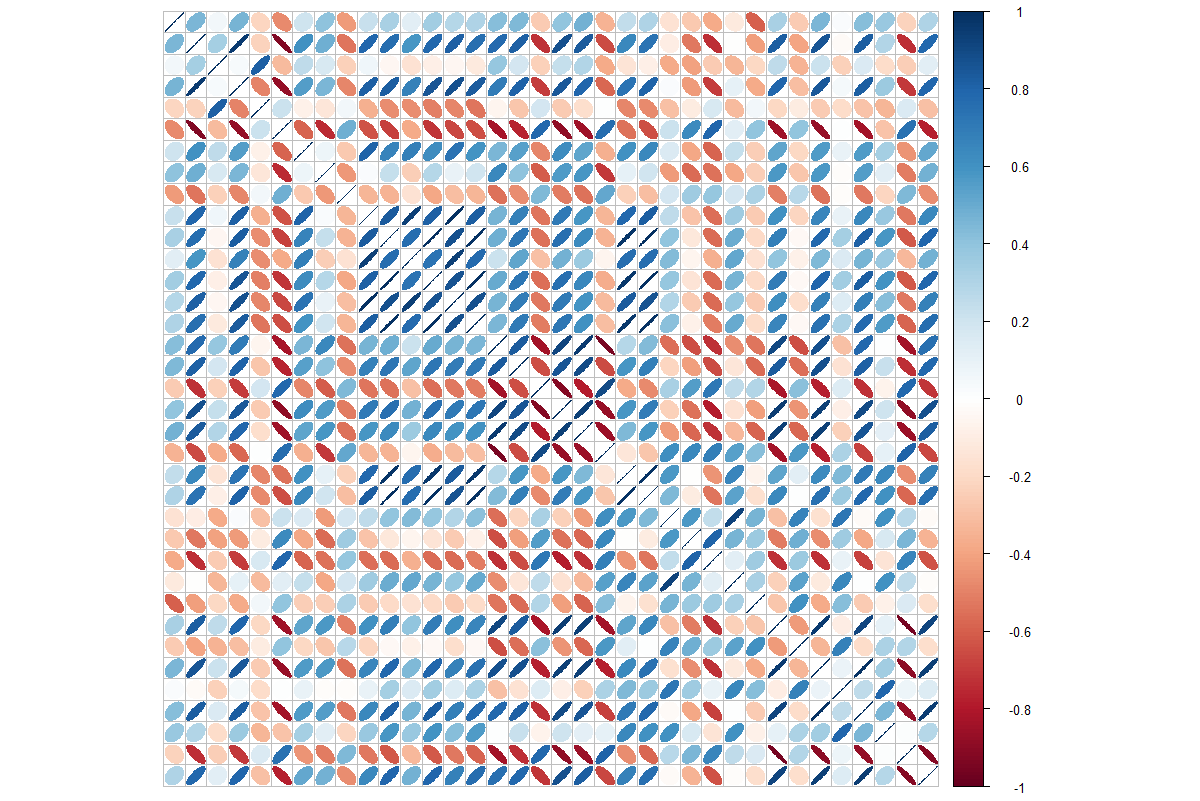}\\
   \caption{Correlation matrix for LED index data-set.}
   \label{fig:corled}
\end{figure}

There are $35$ explanatory variables, $\boldsymbol{\beta}$, $\boldsymbol{\gamma}$, $\boldsymbol{\delta}$ and $\boldsymbol{\zeta}$ are vectors, containing the parameters for each carrier, and $\epsilon$ is an error term assumed to be i.i.d. following a Gaussian distribution. The classical approach to estimate the regression parameters is using Ordinary Least Squares (OLS). The problem here is that the way of acquisition of the data, which is obtaing features from processing images from satellite technology, may imply the presence of atypical observations that could invalidate the results. Therefore, robust methodologies need to be used. On the other hand, the large number of variables derived from the Google Earth image, particularly those of spectral, texture and structure types, are substantially correlated (Figure \ref{fig:corled}).

The multicollinearity issue violates another assumption for using OLS to estimate the parameters of the model. The authors propose to use a dimensionality-reduction step to preserve as much of the variation contained in the entire set of variables while eliminating
collinearity. They performed a principal components analysis (pca) (\cite{Jolliffe2011}, \cite{Ballabio2015}) on all the spectral, texture and structure variables, which makes a total of 27 variables, and after the analysis they propose to use only the first four components because they accounted for $90\%$ of the total variance. Other methods for data containing columns of uninformative variables in the regression problem have been proposed in the literature as well (\cite{Hoffmann2015}, \cite{Li2018}, \cite{Wang2019}). The four extracted components were used as regressors, together with the three land cover variables that prove most relevant: water, shadow, and vegetation. They came up with this result about the relevance by using another approach, but from machine learning area, which is the random forest (RF), since one of the main objectives of the paper was to study the potential of modern machine learning techniques: RF and gradient boost regressor (GBR), in the estimation of socioeconomic indices with remote-sensing data. Focusing on the classical OLS regression, the authors obtained that the third and fourth components were significant, as well as the proportion of an area occupied by water and vegetation.

We propose to study if the results can be improved by using robust regression methods. Let us apply the proposed SR approach and compare it with LTS, S, REWLSE and MM. The raw data, kindly provided by the authors was pre-processed the same way as they propose, by applying pca to the last 27 explanatory variables and join the first four components with the three land cover variables: water, shadow and vegetation, which makes a total of 7 explanatory variables. Table \ref{tab:r2LED} shows the adjusted $R^2$ of the models estimated by each method.

\begin{table}[htbp]
  \centering
  \caption{$R^2$ with (pca transformed) LED index data-set.}
     \resizebox{11cm}{!}{
     \begin{tabular}{r|l|l|l|l|l|l}
          Method & OLS & SR & LTS & S & REWLSE & MM   \\
    \midrule
    $R^2$    &  0.5059 & \textbf{0.6716} & 0.6287 & 0.6031 & 0.5904 & 0.6166 \\
    \end{tabular}%
    }
  \label{tab:r2LED}%
\end{table}%

Variables PC3, PC4, water and vegetation resulted significant in the model obtained by the methods. The percentage of variability explained by the robust methods shows the advantage of robust regression. The $R^2$ of SR is higher than that of the other approaches, although not as high as one would wish. The authors compare the results from OLS with the application of the two machine learning approaches. RF showed an $R^2=0.9354$ and GBR an $R^2=0.8320$. They were interested in finding the best possible model with the ability of capture as much proportion of the variation inherent in the data as possible. But the problem here is the drawback both machine learning methods have in terms of interpretability. Also, as the authors point out, RF and GBR suffer from the issue of overfitting. That is why they propose a cross validation (CV) study. It consisted on dividing the data in two groups, one to train the model, and the other one to test its predictive performance. The 5-fold CV was used and the procedure was repeated 250 times, to obtain the scores for the $R^2$, as in the paper. The scores for the MSE of the response are also saved. Table \ref{r2} shows the median cross-validated $R^2$ obtained by the authors for RF and GBR together with the one we obtained for method SR.

\begin{table}[htbp]
  \centering
  \caption{Median cross-validated $R^2$ with (pca transformed) LED index data-set.}
     \resizebox{5.5cm}{!}{
     \begin{tabular}{r|l|l|l}
          Method &  SR & RF & GBR   \\
    \midrule
   $R^2$    &  \textbf{0.6704} & 0.54 & 0.50  \\
    \end{tabular}%
    }
  \label{r2}%
\end{table}%

\begin{figure}[H]
\centering
\subfigure[Cross-validated $R^2$]{\includegraphics[width=0.45\textwidth]{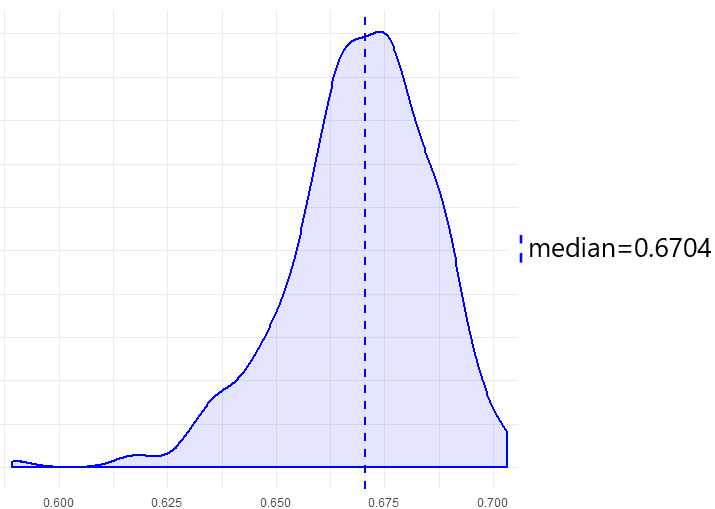}\label{cvr}}
\subfigure[Cross-validated MSE]{\includegraphics[width=0.45\textwidth]{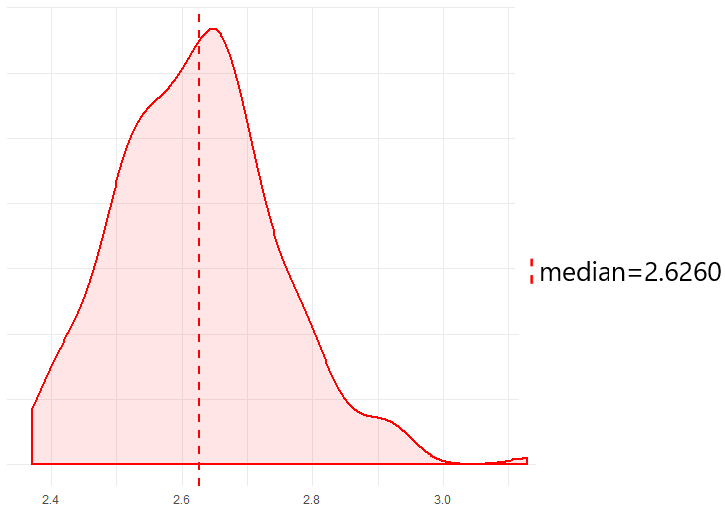}\label{cvm}}
\caption{CV scores and median values (dashed line), with pca.} \label{fig:cv}
\end{figure}

The results show that SR is more robust to overfitting since the $R^2$ is reduced slightly, while that of RF and GBR are significantly reduced. Between the three values, SR has the highest median cross-validated $R^2$. On the other hand, for method SR, the median absolute deviation from the data's median (MAD) of these scores is 0.0145 which is low, meaning that the uncertainty is under control. Figure \ref{cvr} shows the distribution of the cross-validated scores for the $R^2$ obtained with method SR and the median value in a dashed line.

Figure \ref{cvm} shows the results for the MSE. The median of the cross-validated MSE is equal to 2.6260 and the MAD is 0.1199 which are also low values.

Since it was mentioned before, the same pca transformation the authors proposed for the data was made for this research. Now, we propose another transformation that improves the performance according to the results: sparse pca (spca) (\cite{Zou2006}, \cite{Gajjar2017}), which has advantages in case of high correlated variables since it is a kind of variable selection transformation. The spca was made over the 27 variables of the three last groups and the first 10 components were selected since they account for $92.04\%$ of the total variance. These 10 components and the three most relevant land cover variables: water, shadow and vegetation were used to estimate the model.

\begin{figure}[H]
\centering
\subfigure[Cross-validated $R^2$]{\includegraphics[width=0.45\textwidth]{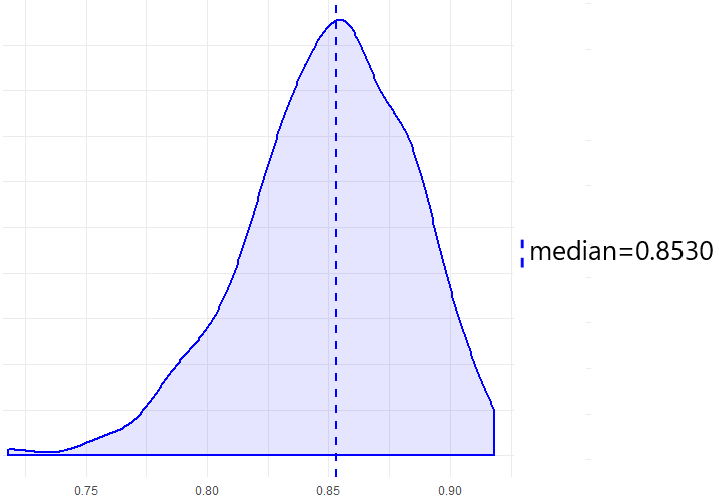}\label{cvrs}}
\subfigure[Cross-validated MSE]{\includegraphics[width=0.45\textwidth]{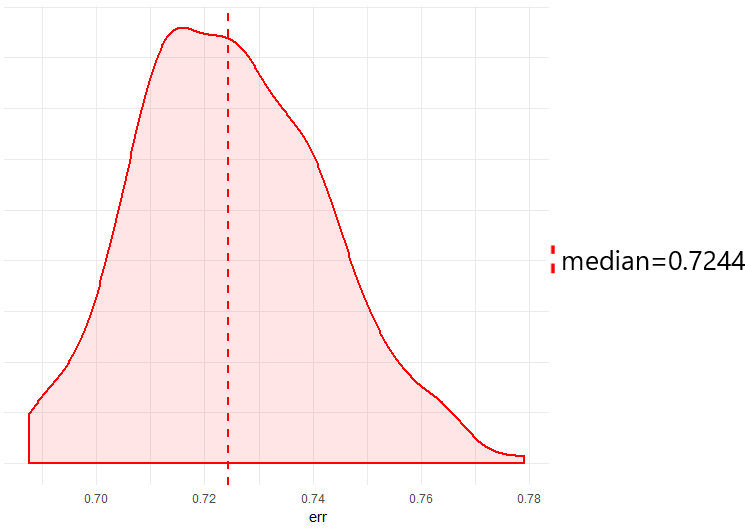}\label{cvms}}
\caption{CV measures and median values (dashed line), with spca.} \label{fig:cvs}
\end{figure}

 Figure \ref{cvrs} shows the distribution of the cross-validated $R^2$ and the median value in a dashed line obtained with SR, which is 0.8530. The MAD of these scores increases to 0.0346 but it is still a low value. Figure \ref{cvms} shows the distribution for the MSE. The median MSE reduces to 0.7244 and the MAD reduces to 0.0177. Table \ref{cvr2spca} shows that the median cross-validated $R^2$ is higher than that obtained with pca transformation but also higher than the obtained with both machine learning techniques, reported in \cite{arribas2017remote}.

\begin{table}[H]
  \centering
  \caption{Median cross-validated $R^2$.}
        \resizebox{7cm}{!}{
        \begin{tabular}{c|c|c|c|c}
          Method &  SR spca &  SR pca & RF & GBR   \\
    \midrule
   $R^2$    &  \textbf{0.8530} & 0.6704 & 0.54 & 0.50  \\
    \end{tabular}%
    }
  \label{cvr2spca}%
\end{table}%

\begin{figure}[H]
 \centering
   \includegraphics[width=0.6\textwidth]{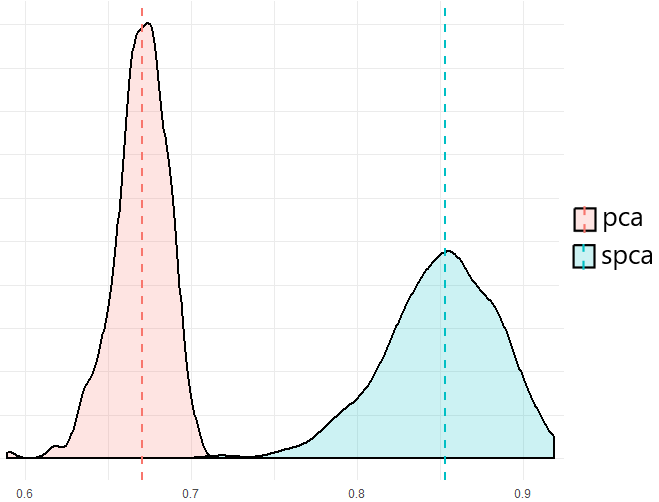}\\
   \caption{Cross-validated $R^2$ and median values (dashed line), for both pca and spca.}
   \label{fig:cvr2both}
\end{figure}

The uncertainty of the obtained $R^2$ is slightly higher with spca transformation, compared to that with the pca transformation. But Figure \ref{fig:cvr2both} shows that the distributions of the $R^2$ scores are quite separated, and the gain is obvious because of the increase in the median value.

Finally, Table \ref{tab:resspca} contains the estimated coefficients, the p-values and the $R^2$ estimated by SR with spca transformation using the complete data-set, which is competitive with respect to the $R^2$ of RF and GBR reported in \cite{arribas2017remote}. As the results point out, the same land cover variables as in the paper remained significant and with the same negative sign, meaning that larger proportions of water and vegetation are associated with smaller deprivation.

\begin{table}[H]
  \centering
  \caption{Results for the model estimated by SR with spca transformation and the $R^2$ for RF and GBR.}
        \resizebox{9cm}{!}{
        \begin{tabular}{lrr|rr}
          & \textbf{coefficient} & \textbf{p-value} &\textbf{RF} & \textbf{GBR}\\
    \midrule
    constant & 0.27191 & 2.03E-05 & & \\
    water   & -1.42641 & 2.00E-16 & & \\
    vegetation  & -0.44513 & 2.00E-05 & & \\
    SPC2  & -0.04409 & 4.51E-03 & & \\
    SPC3  & 0.13215 & 1.52E-06 & & \\
    SPC4  & 0.32566 & 1.03E-15 & & \\
    SPC5  & -0.26745 & 2.35E-11 & & \\
    SPC7  & -0.13735 & 2.24E-03 & & \\
    SPC8  & 0.19544 & 1.64E-03 & & \\
    \midrule
    \textbf{$R^2$} & 0.86820 &  & 0.9354 & 0.8320\\
    \end{tabular}%
    }
  \label{tab:resspca}%
\end{table}%

\section{Conclusions}\label{final}

In the paper, the performance of the proposed SR approach is compared to the classical OLS and other existing robust regression methods. The robust alternatives in the literature have some drawbacks and their performance depend on decisions that, in case of real data, increase the difficulty of robustly estimate the regression parameters. On the other hand, not all available methods have a good behavior in case of large data-sets, high dimension, not all are scalable in terms of computational time, proven to be sufficiently resistant to the presence of outliers. The proposal in this paper is to use the notion of \textit{shrinkage} in order to define robust estimators of location and scatter to estimate the regression parameters. The approach passes through a pair of weighting steps depending on robust Mahalanobis distances, which results in the shrinkage reweighted (SR) regression estimator. The advantages of using the shrinkage are shown in the simulation study and some conclusions can be noted. SR approach yielded competitive results compared to the alternative robust methods from the literature for the regression problem, even in high dimension, heavy-tailed distributed errors, large contamination or transformed data. Furthermore, SR is quite stable computationally since it involves contributions from all the observations instead of sub-sample iterations from the data. Finally, the results with the real data-set examples bear out with the conclusions from the simulation study. Specially with the LED index data where the SR approach provides an improvement of the cross-validated $R^2$ and MSE with respect to classical OLS and machine learning techniques RF and GBR, while maintaining the advantage of interpretability. It remains to be examined as future research if the proposal could be improved by using adjusted quantiles instead of the classical choices from the literature $q_1$ and $q_2$ from Equation \ref{quanti}, which are derived from the chi-squared distribution.

\small

\bibliographystyle{spbasic}
\bibliography{bbib}

\end{document}